\documentclass[nonacm, acmsmall]{acmart}

\usepackage{multirow}
\usepackage{multicol}
\usepackage{longtable}
\usepackage[inline]{enumitem}
\usepackage{hyphenat}
\usepackage{float} 
\usepackage{xcolor}
\newcommand{\ok}{\color{black}}  

\newcommand{\noteAboutOrthogonal}{Variants that can independently be combined with others are marked as \emph{orthogonal}.}

\def\dimensionIconNotation{0}

\if\dimensionIconNotation1
    \usepackage[inkscapeformat=pdf]{svg}
    \newcommand{\headlineAccuracy}{Accuracy}
    \newcommand{\headlineCompleteness}{Completeness}
    \newcommand{\headlineConsistency}{Consistency}
    \newcommand{\headlineCurrentness}{Currentness} 
    \newcommand{\headlineAccessibility}{Accessibility}
    \newcommand{\headlineCompliance}{Compliance}
    \newcommand{\accuracy}{\includesvg[height=0.08in]{images/accuracy1}}
    \newcommand{\completeness}{\includesvg[height=0.08in]{images/completeness}}
    \newcommand{\consistency}{\includesvg[height=0.08in]{images/consistency}}
    \newcommand{\currentness}{\includesvg[height=0.08in]{images/currentness}}
    \newcommand{\accessibility}{\includesvg[height=0.065in]{images/accessibility}}
    \newcommand{\compliance}{\includesvg[height=0.08in]{images/compliance}}
    \newcommand{\valueLevel}{\textbullet}
    \newcommand{\rowLevel}{\includesvg[width=0.08in]{images/row}}
    \newcommand{\columnLevel}{\includesvg[height=0.08in]{images/column}}
    \newcommand{\tableLevel}{\includesvg[height=0.08in]{images/table}}
\else
    \newcommand{\headlineAccuracy}{\textbf{\underline{acc}}uracy}
    \newcommand{\headlineCompleteness}{\textbf{\underline{compl}}eteness}
    \newcommand{\headlineConsistency}{\textbf{\underline{cons}}istency}
    \newcommand{\headlineCurrentness}{\textbf{\underline{cur}}rentness}
    \newcommand{\headlineAccessibility}{\textbf{\underline{ac}}ce\textbf{\underline{s}}si\textbf{\underline{b}}ility}
    \newcommand{\headlineCompliance}{\textbf{\underline{c}}o\textbf{\underline{mpli}}ance}
    \newcommand{\accuracy}{acc}
    \newcommand{\completeness}{compl}
    \newcommand{\consistency}{cons}
    \newcommand{\currentness}{cur}
    \newcommand{\accessibility}{acsb}
    \newcommand{\compliance}{cmpli}
    \newcommand{\valueLevel}{V}
    \newcommand{\rowLevel}{R}
    \newcommand{\columnLevel}{C}
    \newcommand{\tableLevel}{T}
\fi

\newcommand{\deequ}{D}
\newcommand{\dbtCore}{d}
\newcommand{\mobyDQ}{M}
\newcommand{\gx}{X}
\newcommand{\sodaCore}{S}
\newcommand{\griffin}{G}
\newcommand{\evidently}{E}

\newcommand{\deequName}{Deequ}
\newcommand{\dbtCoreName}{dbt Core}
\newcommand{\gxName}{GX}

\newcommand{\griffinName}{Griffin}
\newcommand{\evidentlyName}{Evidently}
\newcommand{\mobyDQName}{MobyDQ}

\newcommand{\categoryColumnWidth}{0.13\textwidth}
\newcommand{\functionalityColumnWidth}{0.3\textwidth}
\newcommand{\functionalitiesTablesStretch}{\def\arraystretch{1.3}}  
\usepackage{soul} 

\newcommand{\numTools}{seven }

\newcommand\twodigits[1]{%
   \ifnum#1<10 0#1\else #1\fi
}

\newcommand\functionalityId[1]{\textit{f-{\twodigits{#1}}.}}

\newcommand\variantId[2]{\textit{f-{\twodigits{#1}}$_{#2}$}.}

\newcounter{functionalityIdCounter}
\newcommand\nextFunctionalityId{\stepcounter{functionalityIdCounter}\functionalityId{\thefunctionalityIdCounter}}

\newcommand\resetFunctionalityId{\setcounter{functionalityIdCounter}{0}}

\newcounter{variantIdCounter}[functionalityIdCounter]
\newcommand\nextvariantId{\stepcounter{variantIdCounter}\variantId{\thefunctionalityIdCounter}{\alph{variantIdCounter}}}

\newcounter{functionality}[section] 

\newenvironment{functionality}[1] 
{%
    \refstepcounter{functionality} 
    \noindent\textit{f-\twodigits{\arabic{functionality}}.} #1 
    \par\noindent%
}
{\par\vspace{0.5em}}

\usepackage[raggedrightboxes]{ragged2e} 

\begin{document}
\title{Unfolding Data Quality Dimensions in Practice: A Survey}

\author{Vasileios Papastergios}
\affiliation{
    \institution{Aristotle University}
    \city{Thessaloniki}
    \state{Greece}
}
\email{papastva@csd.auth.gr}

\author{Lisa Ehrlinger}
\affiliation{
    \institution{Hasso Plattner Institute, University of Potsdam}
    \city{Potsdam}
    \state{Germany}
}
\email{lisa.ehrlinger@hpi.de}

\author{Anastasios Gounaris}
\affiliation{
    \institution{Aristotle University}
    \city{Thessaloniki}
    \state{Greece}
}
\email{gounaria@csd.auth.gr}

\begin{abstract}
    Data quality describes the degree to which data meet specific requirements and are fit for use by humans and/or downstream tasks (e.g., artificial intelligence). Data quality can be assessed across multiple high-level concepts called \emph{dimensions},
    such as accuracy, completeness, consistency, or timeliness.
    While extensive research and several attempts for standardization (e.g., ISO/IEC 25012) exist for data quality dimensions, their practical application often remains unclear.
    In parallel to research endeavors, a large number of tools have been developed that implement functionalities for the detection and mitigation of specific data quality issues, such as missing values or outliers.
    With this paper, we aim to bridge this gap between data quality theory and practice by systematically connecting low-level functionalities offered by data quality tools with high-level dimensions, revealing their many-to-many relationships.
    Through an examination of \numTools open-source data quality tools, we provide a comprehensive mapping between their functionalities and the data quality dimensions, demonstrating how individual functionalities and their variants partially contribute to the assessment of single dimensions.  
    This systematic survey provides both practitioners and researchers with a unified view on the fragmented landscape of data quality checks, offering actionable insights for quality assessment across multiple dimensions. 
\end{abstract}

\maketitle

\section{Introduction}
\label{section:Introduction}

In today's data-driven era, the importance of data quality (DQ) has become critical~\cite{Coleman_2013, Batini2016, Nagle_2020, Lee_2014}. 
Organizations increasingly rely on exponentially growing volumes of data to drive operational, tactical, and strategic decisions, either directly or through training machine/deep learning models. Consequently, the quality of these data plays a crucial role in downstream tasks across the whole operational pipeline of modern businesses and organizations. \emph{High-quality} data are essential for ensuring the integrity of analytics~\cite{Jugulum2016_importance_dq_analytics}, empowering machine learning (ML) models~\cite{rein_edbt_2023, lee2021_survey_cleaning_methods_ML, Li2019CleanMLAB, budach2022effects} and supporting business intelligence (BI) efforts~\cite{hartl_dq_business_intelligence,torres_information_quality_bi}. Poor data quality can lead to costly, misguided decisions, operational inefficiencies, and a loss of trust among stakeholders or end users~\cite{Rana2021_operational_inefficiencies,Loshin2011_business_impacts_of_dq,Redman1998_impact_poor_dq_typical_enterprise, budach2022effects,breck_data_validation_machine_learning}. 
Managing DQ effectively becomes even more pivotal with the advent of generative artificial intelligence~\cite{Davenport_2024,mohammed2024qualityAssessmentVision, Wang_data_quality_Gen_AI_2023, Wrsdrfer_data_quality_ai_2023}.

Over the last decades, extensive research has been conducted on how to \textit{assess} and \textit{improve} data quality in terms of methodologies~\cite{Batini_2009,Cichy_2019}, alongside public standards, such as the ISO/IEC 25012 standard~\cite{iso25012}. The literature widely acknowledges that DQ is characterized by multiple \textit{DQ dimensions}, also known as \emph{characteristics} or \emph{attributes}~\cite{Wang_1998,Batini2016}, such as, accuracy, completeness, or timeliness. These dimensions can be quantified with \textit{DQ metrics}, which are functions that map a DQ dimension to a numerical value. Despite extensive discussions on DQ dimensions in literature, their practical implementation often remains unclear.

In parallel to research endeavors, a wide variety of DQ tools have been developed~\cite{survey_ehringer}. These tools are software products that provide practical solutions, assisting businesses and organizations to assess and improve their data quality. 
Although a great amount of implementation effort has been devoted to these tools, significant heterogeneity exists in (i) the terminology used to describe the functionalities in these DQ tools, as well as (ii) the alleged - and often overlooked - connection of the functionalities with the DQ dimensions. 
For example, checking that all rows in a table satisfy a certain constraint can be found as \emph{conformance}, \emph{compliance}, or even \emph{validity} in the tools, while all referring to the same software engineering functionality.
Conversely, the same term may denote different functionalities across tools. \emph{Completeness}, for example, can be found referring to both functionalities that count the number of rows in a table and functionalities that check for the presence of NULL values. One reason for this fragmented landscape could be the lack of consistent terminology within the DQ dimensions themselves. Despite available standards, terms like \emph{currentness}, \emph{freshness}, \emph{recency}, and \emph{timeliness} are sometimes used interchangeably~\cite{Batini2016} and even the understanding of \emph{accuracy} is not unified~\cite{Haegemans_2016}.
The heterogeneous terminology and consequently inconsistent use of both DQ dimensions and concrete functionalities implemented in DQ tools implies two major challenges: 
\begin{description}
    \item[(C1)] Practitioners are left with the question on how to use DQ dimensions in practice and which DQ functionalities can be expected by which tool. 
    \item[(C2)] Researchers are left with the question how specific DQ dimensions are materialized in practice. 
\end{description}

To address these challenges, this paper aims to connect the dots between DQ theory and practical implementations. 
Note that we focus on DQ \emph{assessment} (i.e., checks that measure specific properties of the data and compare it with a threshold or reference data) and do not consider functionalities for DQ \emph{improvement} (i.e., data cleaning).
We investigated the \emph{low-level}\footnote{We use the term \emph{low-level} functionalities to describe concrete data verification checks and to differentiate our work from surveys that examine more general concepts such as the presence of data profiling functionalities in tools.} functionalities implemented in \numTools widely used open-source DQ tools to reflect the practical perspective. 
In addition, we use the ISO/IEC 25012 standard~\cite{iso25012}, which defines a DQ model with 15 \emph{high-level} DQ dimensions (characteristics), as listed in \autoref{appendix:dimensions}, to reflect the theoretical perspective. Our contributions can be summarized as follows:\footnote{A preliminary short version of our work has appeared in \cite{DQarxiv} as a technical report.}

\begin{enumerate}
    \item We provide a unifying list of low-level functionalities offered by  widely used open-source DQ tools.
    \item We present detailed catalogs of variants per low-level functionality. These lists aim to offer actionable guidelines to DQ practitioners, providing answers regarding \emph{what} error detection checks can be performed on the data at hand and \emph{how} these checks are implemented at the \emph{source code} level by widely used open-source tools.
    \item Taking a step further, we introduce a novel mapping between the low-level functionalities identified in the DQ tools and the DQ dimensions from the ISO/IEC 25012 standard. The mapping provides both researchers and practitioners with a unified view on DQ assessment and reveals the underlying connection between successfully used error detection checks  in DQ tools and theoretical DQ dimensions. 
\end{enumerate}

The rest of this paper is structured as follows. We present the methodology to conduct this survey in~\autoref{section:methodology} and the results in form of the list of low-level functionalities and their mapping to DQ dimensions in~\autoref{section:survey_findings}. For each low-level functionality, a list of implementation variants is provided in Sections~\ref{subsection:value_checks} to~\ref{subsection:dataset_checks}. 
In \autoref{section:materialization_dimensions}, we reexamine our findings through the lens of DQ dimensions, highlighting key observations about their materialization in practice. 
Related work is discussed in \autoref{section:related_work} and we conclude this paper with an outlook on future work in \autoref{section:conclusion_future_work}.

\section{Methodology}
\label{section:methodology}

\begin{figure*}[tb!]
    \centering
    \includegraphics[width=\linewidth]{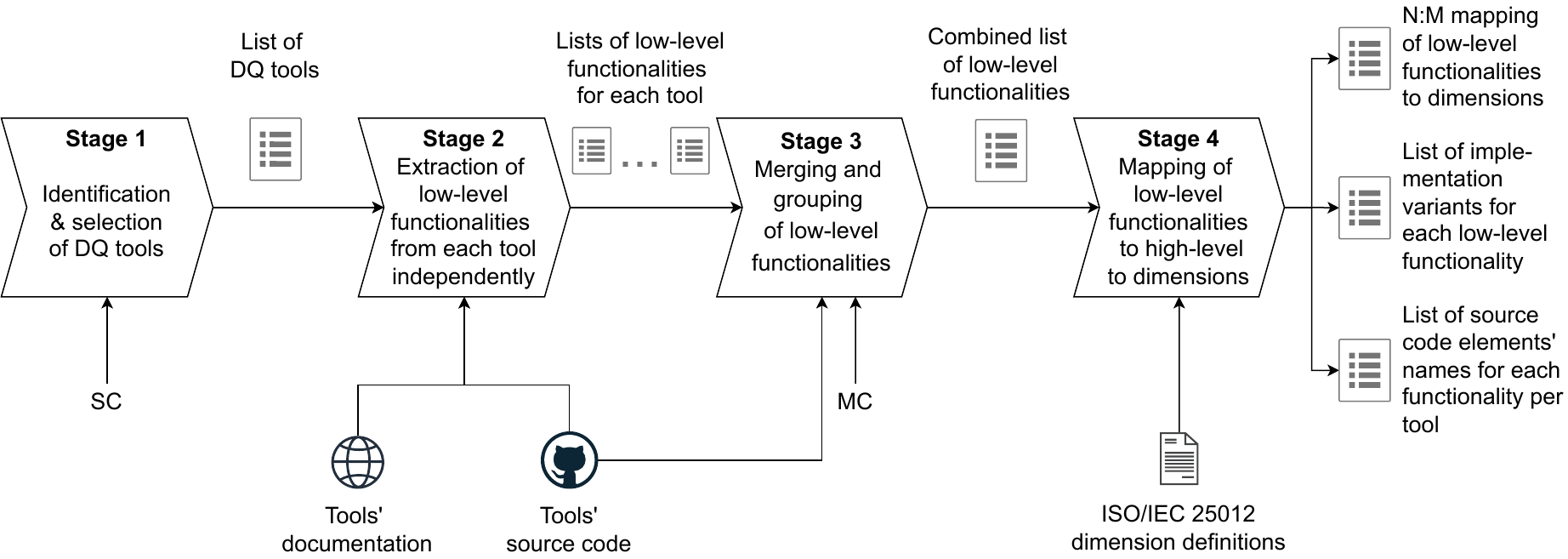}
    \caption{The methodology used to conduct the survey.}
    \Description{A diagram with the steps adopted to conduct the survey.} 
    \label{img:methodology}
\end{figure*}

In this section, we explain our approach for conducting the survey, which consists of the following four stages:
\begin{description}
    \item[Stage 1:] Identification and selection of DQ tools.
    \item[Stage 2:] Extraction of low-level functionalities from each tool independently.
    \item[Stage 3:] Merging and grouping of low-level functionalities.
    \item[Stage 4:] Mapping of low-level functionalities to high-level dimensions.
\end{description}
\autoref{img:methodology} illustrates the input and output of each stage, and their integration into a complete workflow. We elaborate on each stage in the rest of this section.

\subsection{Identification and selection of DQ tools}
The foundation of our survey lies in identifying and selecting representative DQ tools based on the following predefined selection criteria~(SC):

\begin{description}[leftmargin=25pt] 
    \item[SC1:] The tool offers DQ management functionalities and is not tied to a specific task (e.g., duplicate detection only). 
    \item[SC2:] The tool is an open source project and its source code is available on a publicly accessible platform (e.g., GitHub).
    \item[SC3:] The tool is actively maintained and/or extended.
    \item[SC4:] The tool is widely used (i.e., not a complete niche product). 
\end{description}

Based on a comprehensive list of previously investigated DQ tools   (cf.~\cite{survey_ehringer,Pulla_DQSurvey_2016,Pushkarev_DQSurvey_2010,Barateiro_DQSurvey_2005}), the well-known Gartner Magic Quadrant for Data Quality Solutions~\cite{GartnerDQTools_2022}, and additional online sources\footnote{ https://atlan.com/open-source-data-quality-tools/}, we compiled the following list of \numTools DQ tools (T) that meet our criteria:
\begin{description}[leftmargin=17pt]
    \item[T1:] \textbf{\dbtCoreName}{\footnote{https://github.com/dbt-labs/dbt-core}}, an SQL-based Python framework that offers a wide range of built-in DQ checks and integrates with many data platforms. It also enables the user to define custom template checks that can be reused with different parameters each time.
    \item[T2:] \textbf{\deequName}{\footnote{https://github.com/awslabs/deequ}}, a Scala library developed on top of Apache Spark  for defining and executing DQ checks on large datasets~\cite{deequ_paper_2018}. The tool also features a Python API (PyDeequ) and rule-based suggestions for DQ checks, based on data profiling.
    \item[T3:] \textbf{\evidentlyName}{\footnote{https://github.com/evidentlyai/evidently}}, a Python framework focused on evaluating and monitoring the performance of AI systems, including large language models. It offers a wide range of built-in DQ metrics and result visualizations.
    \item[T4:] \textbf{Great Expectations (\gxName)}{\footnote{https://github.com/great-expectations/great\_expectations}}, a Python library that enables users to define \emph{expectations}, i.e., verifiable assertions about data. It combines built-in expectations with community-contributed checks.
    \item[T5:] \textbf{\griffinName}{\footnote{https://github.com/apache/griffin}}, {an Apache Software Foundation framework} for DQ monitoring of distributed data, built on Apache Hadoop (in Java) and Spark (in Scala). It enables the user to define DQ checks in several built-in categories and to visualize the results.
    \item[T6:] \textbf{\mobyDQName}{\footnote{https://github.com/ubisoft/mobydq}}, a desktop application offering built-in DQ \emph{indicators}, i.e., metrics that describe the quality of data. It combines a Python-based backend with a graphical interface to define and execute DQ checks.
    \item[T7:] \textbf{Soda Core}{\footnote{https://github.com/sodadata/soda-core}}, a Python library for measuring and monitoring DQ on a variety of data platforms. The checks are written in a domain-specific language, porting a low-code approach that enables both technical and non-technical collaborators within organizations to define checks.
\end{description}

\subsection{Independent extraction of low-level functionalities}

Following tool selection, we conducted a thorough examination of each tool's official documentation and source code, accessed through their websites and GitHub repositories respectively.
Our target for this stage was to extract the low-level functionalities offered by each tool \emph{independently}. This implies producing a separate list of functionalities per DQ tool, i.e., \numTools different lists, regardless of exact or near duplicate functionalities among the tools. No grouping or merging of functionalities was performed at this stage. 

The decision to let the tool's implementations guide our analysis without having a predefined list of expected functionalities (as for example done in~\cite{survey_ehringer}) was a strategic one.
This approach allowed to collect any low-level functionalities from the tools from an \emph{unbiased perspective}.
The outcome comprised \numTools lists of low-level functionalities, each annotated with the specific terminology used within that tool.

\subsection{Merging and grouping of low-level functionalities}
    Here, we focused on identifying and reconciling exact and near duplicate functionalities among DQ tools that were intentionally overlooked in the previous stage. The goal was to combine the separate lists into a single comprehensive list meeting the following merging criteria (MC):
        \begin{description}[leftmargin=28pt]
            \item[MC1:] Lossless merging: every low-level functionality present in at least one tool must be represented in the combined list.
            \item[MC2:] Concise representation: each distinct low-level functionality should appear exactly once in the combined list, in sufficient depth but without redundancy.
        \end{description}

        Meeting these criteria proved challenging, primarily due to the diverse and sometimes conflicting terminology across tools. To overcome this challenge, we conducted detailed source code examinations, trying to focus on the core functionality. The heterogeneity in programming languages and architectures across tools required systematic effort to thoroughly understand both the implementation details and the underlying software engineering decisions made by the developers. We provide insights into such choices wherever applicable in the rest of our work.
    The main actions performed in this stage are \emph{merging} and \emph{grouping} of low-level functionalities. More specifically, during the compilation of the combined list:
        \begin{itemize}
            \item we \textbf{merge} functionalities when their implementations indicates that they are exact or near duplicates. For near duplicates, we consider one functionality as an \emph{implementation variant} of the other, as detailed in~\autoref{section:survey_findings}.
            \item we \textbf{group} functionalities that are closely semantically related (e.g., they are distribution-based). The novel classification scheme we have adopted is based on the functionality semantics, as explained in \autoref{subsection:presentation_functionality_categories}.
            The grouping phase starts after no further merging of functionalities can be performed. Its main target is to ease the presentation of the functionalities, adding an orthogonal viewpoint to the analysis of their spectrum.
        \end{itemize}
    The outcome of this stage is a comprehensive list of low-level functionalities that \emph{sufficiently} and \emph{concisely} cover the spectrum of low-level functionalities extracted separately from the tools.

\subsection{Mapping of low-level functionalities to high-level dimensions}
This stage aims to systematically connect the identified low-level functionalities with 6 selected ISO/IEC 25012 high-level DQ dimensions. 
The selection was made  on the grounds that these dimensions are mainly targeted by the aforementioned tools in an application-agnostic manner. These DQ dimensions broadly represent \emph{inherent} data quality according to the standard's terminology, 
i.e., the intrinsic potential of data to meet defined needs. 
The connection between low-level functionalities and high-level DQ dimensions is neither trivial, nor straightforward. It turns out that the mapping between real-world implementations and DQ dimensions is an N:M (many-to-many) relation, as detailed in~\autoref{section:survey_findings}. The main challenge of this stage was to, (1) conceptualize and, (2) to apply a \emph{systematic} approach for identifying the most closely related DQ dimensions for each functionality. We base our analysis on the definitions from ISO/IEC 25012 (listed in~\autoref{appendix:dimensions}). We establish a connection when the purpose of a functionality directly or indirectly aligns with the textual definition of a DQ dimension. Detailed justifications for these connections are provided in~\autoref{section:survey_findings}.

\section{Survey Findings}
\label{section:survey_findings}

\begin{table*}[tb!]
    \functionalitiesTablesStretch
    \caption{A comprehensive list of the low-level functionalities (column 2) extracted from the investigated DQ tools, grouped into categories (column 1). {\ok Each low-level functionality is associated with the granularity level(s) it operates at (columns 3-6), its closely related dimension(s) (columns 7-12) and the tools that implement it (columns 13-19). In case multiple granularity levels are found, the level at which the functionality \emph{mainly} operates is marked with an asterisk.}}
    \label{table_functionalities_to_dimensions}
    \resizebox{\linewidth}{!}{%
\begin{tabular}{ccccll|cccccc|ccccccc}
            \toprule
                                                                                                                                                  & \multicolumn{1}{c|}{\multirow{7}{*}{\textbf{Low-level functionality}}}                                                                                                            & \multicolumn{4}{c|}{\textbf{Granularity}}                             & \multicolumn{6}{c|}{\textbf{DQ dimensions}}                         & \multicolumn{7}{c}{\textbf{DQ tools}}                                                                                                                                                                                                                                                                                                                                                                                                                                                                                                                                                                                                                                                                                                                                                                                                                                                                                                                                                                                                                                                                                                                                                \\
\bottomrule
        \end{tabular}
    }
\end{table*}
\resetFunctionalityId

\autoref{table_functionalities_to_dimensions} summarizes the main findings of our survey: a comprehensive mapping between low-level functionalities and high-level DQ dimensions. The \textbf{second column} lists all \textit{low-level functionalities} (i.e., error detection data checks\footnote{In our work, we use the terms low-level functionalities to describe specific data verification checks, which contribute to DQ assessment but do not improve DQ. })
identified in the examined tools. Each low-level functionality is linked to its detailed discussion in the corresponding Subsection for easy navigation. The functionalities are grouped into broader \textit{categories}, shown in the \textbf{first column}, facilitating the joint presentation of related functionalities. However, our primary focus remains on revealing the connection between \emph{functionalities}, rather than \emph{categories}, and DQ dimensions, as functionalities within the same category may relate to different dimensions. 

The next four columns (\textbf{columns 3-6}) represent the data granularity level(s) (e.g., row, column) at which each functionality operates. Our novel classification scheme enables us to capture functionalities that can operate at multiple levels, as explained below. The following six columns (\textbf{columns 7-12}) represent the selected ISO/IEC 25012 dimensions that are affected by these low-level functionalities. Out of the fifteen dimensions defined by the standard, we were able to identify six that relate to the recorded functionalities, broadly, those addressing \emph{inherent} data quality.

The remaining \numTools columns (\textbf{columns 13-19}) represent the examined DQ tools, indicating whether a tool \emph{directly} implements a functionality through built-in checks. Note that unmarked functionalities might still be achievable through indirect or user-defined way.
Our survey's objective is \emph{not} to compare the tools. Instead, we treat these tools as representatives of the current error detection landscape, through their built-in low-level functionalities.

A comprehensive list of the \emph{exact names} by which the functionalities are found in the tools is presented in~\autoref{appendix:function_names}. In the rest of the section, we elaborate on each identified functionality, beginning with an overview of their categories.

\subsection{Categories of DQ low-level functionalities}
\label{subsection:presentation_functionality_categories}

This section describes the categories of low-level functionalities in \autoref{table_functionalities_to_dimensions}. We introduce these categories to group semantically related low-level functionalities and help readers comprehend the content.

Our classification comprises five categories: 
\begin{enumerate*}
    \item Conformance checks;
    \item Distribution-based checks;
    \item Volume- \& Cardinality-based checks;
    \item Correlation-based checks; and
    \item ML-oriented checks.
\end{enumerate*}
Conformance checks are further subdivided in~\autoref{table_functionalities_to_dimensions} based on their operational scope: single value, row, column, table schema, or entire table, respectively. 

The novel classification scheme we have adopted decouples  the nature (i.e., semantics) of the observed functionalities from the data granularity level (e.g., row, column) at which they operate as much as possible. Such a choice enables us to capture low-level functionalities that, along with their \emph{implementation variants}, cover a broad and complicated spectrum, spanning across multiple granularity levels. This unifying categorization based on the \emph{semantics} of the low-level functionalities can provide the reader with actionable insights into the different families of error detection checks they can apply in their own use cases.

Notably, our classification criterion does not explicitly consider whether functionalities operates directly on the data values or on metadata, either inherent (e.g., schema) or derived (e.g., distinct value counts). Regarding derived metadata, it is widely accepted that \emph{data profiling} is integral to successful DQ assessment~\cite{survey_ehringer, azeroual_profiling_2018}. Indeed, our findings reveal a numerous low-level functionalities operating on profiling results, rather than raw data values. These functionalities appear across multiple categories, suggesting that this distinction represents an orthogonal classification dimension. We incorporate this perspective when analyzing each category's functionalities in our subsequent discussions, where we shift our focus towards the mapping of low-level functionalities to high-level DQ dimensions.

\subsection{Conformance Checks at the Value Level}
\label{subsection:value_checks}
This category includes low-level functionalities for error detection that address the question \emph{``What checks can be executed on single data values (i.e., on the cell level) to assess their quality?''} Most checks produce a binary output (\emph{pass} or \emph{fail}) for each input value, determinable by examining the value in isolation. We explicitly note any exceptions to this pattern where applicable.

A common feature among the examined DQ tools is their support for column- or table-level aggregations. A \emph{column-level} aggregation computes the number or fraction of column values for which the check is successful, over the total number of column values. A \emph{table-level} aggregation identifies rows for which the check fails, optionally presenting a sample for manual inspection.

Table-level aggregations are typically implemented using SQL \texttt{SELECT} queries (or equivalents, e.g., \texttt{pandas.DataFrame} operations in Python). The \texttt{WHERE} clause contains the negation of the value-level check, thereby identifying failing rows. When only a sample of the failing rows is required, most tools use simple \texttt{LIMIT} clauses, though some implement more sophisticated sampling strategies. Column-level aggregations are readily achieved by adding \texttt{COUNT} functions on top of these queries. 
In total, value checks can be summarized as follows:

\begin{description}
    \item[Input:] a single data value (i.e., a data cell)
    \item[Output:] pass/fail
    \item[Add-ons:] column-level aggregation, table-level aggregation
    \item[Source code:] simple SQL \texttt{SELECT}-\texttt{WHERE}-\texttt{COUNT} queries or equivalent
\end{description}
\autoref{table_functionalities_value} presents the low-level functionalities in this category with their implementation variants, as identified across the examined tools. 

\begin{table*}[tb!]
    \caption{The low-level functionalities in the category of value checks with their implementation variants found in the tools examined.~\noteAboutOrthogonal}
    \label{table_functionalities_value}
    \footnotesize
    \functionalitiesTablesStretch
    \begin{tabular}{ll} 
        \toprule
        \textbf{Low-level functionality}                                                                    & \textbf{Implementation variants}                                                    \\
        \midrule
        \multirow[t]{9}{\functionalityColumnWidth}{\nextFunctionalityId~ values fall within range or set}   & \nextvariantId~ range of allowed values with inclusive/exclusive endpoints          \\
                                                                                                            & \nextvariantId~ dynamic range endpoints with row-level scope                        \\
                                                                                                            & \nextvariantId~ set of allowed values                                               \\
                                                                                                            & \nextvariantId~ set of allowed pairs with row-level scope                           \\
                                                                                                            & \nextvariantId~ set of \emph{not} allowed values                                    \\
                                                                                                            & \nextvariantId~ referential integrity                                               \\
                                                                                                            & \nextvariantId~ only \emph{distinct} values are checked (orthogonal)                \\
                                                                                                            & \nextvariantId~ execution of \texttt{WHERE} clause before the check (orthogonal)    \\
                                                                                                            & \nextvariantId~ only the \emph{most frequent} value is checked (orthogonal)         \\
        \midrule
        \multirow[t]{3}{\functionalityColumnWidth}{\nextFunctionalityId~ string values are of right length} & \nextvariantId~ equality to a desired value                                         \\
                                                                                                            & \nextvariantId~ threshold-based comparisons                                         \\
                                                                                                            & \nextvariantId~ simple statistics on values' lengths                                \\
        \midrule
        \multirow[t]{7}{\functionalityColumnWidth}{\nextFunctionalityId~ values comply with regex}          & \nextvariantId~ user-defined regex                                                  \\
                                                                                                            & \nextvariantId~ built-in regex for common cases, e.g., email, URL, etc.             \\ 
                                                                                                            & \nextvariantId~ regex to \emph{not} match                                           \\
                                                                                                            & \nextvariantId~ list of allowed/disallowed regexes                                  \\
                                                                                                            & \nextvariantId~ comparison against reference data (orthogonal)                      \\
        \midrule
        \multirow[t]{1}{\functionalityColumnWidth}{\nextFunctionalityId~ timestamp values are recent}       & \nextvariantId~ recentness based on current system time (freshness)                 \\
                                                                                                            & \nextvariantId~ recentness based on reference data (latency)                        \\
                                                                                                            & \nextvariantId~ execution of \texttt{WHERE} clause before the check (orthogonal)    \\
                                                                                                            & \nextvariantId~ execution of \texttt{GROUP BY} clause before the check (orthogonal) \\
        \bottomrule
    \end{tabular}
\end{table*}


\subsubsection{Values fall within range or set~\ref{func:value:range_set}}
\label{subsubsection:value:range_set}
This low-level functionality verifies that values either fall within a user-defined range or belong to a set of accepted values. For instance, it can ensure that an employee count for a project stays within bounds (e.g., non-negative and not exceeding fifty for a fifty-person company), or that a payment transaction status is strictly one of ``Completed'', ``Declined'', or ``Pending''.

\paragraph{Variants}
For \emph{range}-based checks, endpoints can be either inclusive or exclusive. They can also be dynamically determined with a row-level scope. Consider a table storing product order data: each row contains both shipped and returned (defected) product counts for a specific order. Rather than applying fixed range endpoints across all orders, dynamic endpoints allow meaningful constraints-for example, ensuring the defective count never exceeds the shipped count for each specific order.

For \emph{set}-based checks, variants include checking value \emph{pairs} with row-level scope against accepted set of pairs, or alternatively, defining sets of \emph{not} allowed values. The latter proves particularly useful in several scenarios.
Consider an organization's document sharing system where each document has a security classification. To prevent unauthorized access, views for ``intern'' roles could explicitly exclude documents classified as ``secret'' or ``top-secret''-especially valuable when dealing with varying classification protocols or multiple data sources. Another variant ensures referential integrity, where allowed values are sourced from another column, typically in a different table.

The orthogonal variants of this functionality include column-level aggregations or row filtering before the actual check. Checking only distinct values can improve efficiency, particularly when distinct value retrieval can be performed in sublinear time w.r.t. the number of checked values (e.g., they are indexed or precomputed). Applying a \texttt{WHERE} clause pre-filters rows before the check, so that the error detection runs only on a subset of the initial data. Focusing on just the most frequent value can reduce the computational resources required, while providing a relaxed form of compliance checking.

\paragraph{Mapping to dimensions} All variants primarily relate to \emph{accuracy}, since out-of-range or disallowed values misrepresent real-world attributes. \emph{Compliance} is highly relevant, due to the existence of some physical restriction (e.g., fifty employees in total) or business rule (e.g., agreed transaction statuses), defining acceptable values. The referential integrity variant additionally connects to \emph{consistency}, ensuring coherence with other data, i.e., the set of accepted references.

\subsubsection{String values are of right length~\ref{func:value:string_length}}
\label{subsubsection:value:string_length}
This low-level functionality verifies user-defined string length constraints. For example, ensuring phone numbers contain exactly 10 digits. Note that this check operates on a derived attribute (string length) rather than the actual value content.

\paragraph{Variants} 
Beyond exact length matching, tools support threshold-based comparisons, validating that the lengths of data values are smaller/greater than a threshold or in between. The most generic variant lets the user define a callback function that receives the length of the checked value and returns a binary output (i.e., pass or fail), encapsulating the logic that decides whether a value length is valid or not. Another variant performs checks on length-based descriptive statistics, which is more relevant to the discussion about statistic-based checks (cf. ~\autoref{subsubsection:column:statistics}).  

\paragraph{Mapping to dimensions} String values of incorrect length may fail to accurately represent the modeled domain, connecting this functionality to \emph{accuracy} both syntactically and semantically. The functionality also relates to \emph{compliance} when length restrictions stem from physical constraints or business rules; for example, security protocols that require all employees having a strong password of 10+ characters for system access.

\subsubsection{Values comply with regex~\ref{func:value:regex}}
\label{subsubsection:value:regex}
This low-level functionality verifies whether data values match specified regular expressions or patterns. For instance, it can ensure product identifiers follow a specific format-starting with ``p'', followed by 6 digits, and ending with 3 capital letters.  

\paragraph{Variants} Beyond user-defined expressions, several tools provide built-in patterns for common cases (e.g., credit card numbers, email addresses, URLs). While these pattens are predefined by tool developers, they ultimately resolve to regular expressions at the source code level. Another variant supports negative matching, ensuring patterns are \emph{not} matched. Consider products available only for in-store purchase ending with the letter ``X''-this variant could ensure that customers browsing products at the firm's website are not presented with such products, since they are not expected to appear in online catalogs. A more generic variant accepts \emph{lists} of allowed or disallowed patterns. This provides syntactic sugar, abstracting the complexity of composite regular expressions with multiple disjunctions.

An orthogonal variant enables comparison against reference data (also found in literature as \emph{gold standard}). This computes matched/mismatched value counts for corresponding columns in both current and reference tables, allowing threshold based comparisons of the differences. This effectively quantifies how closely the data adhere to patterns observed in ideal or desired data.

\paragraph{Mapping to dimensions} This functionality primarily connects to \emph{accuracy}, as pattern-mismatched values often misrepresent reality-for example, a malformed email address missing a ``@'' symbol can never correspond to an actual person. \emph{Compliance} becomes relevant when patterns enforce business rules, such as specific product identifier formats.

\subsubsection{Timestamp values are recent~\ref{func:value:timestamp}}
\label{subsubsection:value:timestamp}
This low-level functionality assesses whether timestamp values are sufficiently recent for the intended use. 
Consider a real-time routing service; data must reflect conditions within the last few minutes to provide actionable insights for drivers. People usually do not care about how busy a road segment was an hour ago, when they are late for an appointment, even if the information is completely accurate.

\paragraph{Variants} Recency assessment follows two main approaches. The first, termed \emph{freshness}, compares recency against the current system time -- broadly what the user understands as \emph{now}. This suits real-time applications like the traffic routing example. The second, termed \emph{latency}, compares recency against reference data timestamps. This variant quantifies data flow delays between source (reference) and destination (data at hand) tables by comparing their latest timestamps, regardless of the current time.

Two orthogonal variants were found for this low-level functionality. Both of them include the execution of some operation before the actual recency check. The first filters several rows of the table at hand by applying an SQL \texttt{WHERE} clause, while the second executes a \texttt{GROUP BY} clause before the check. 

\paragraph{Mapping to dimensions} This functionality primarily relates to \emph{currentness}, since it directly assesses whether data are of the right age. The \emph{context} (e.g., task, system, humans involved, etc.)~\cite{fadlallah_data_quality_context_2023, serra_data_quality_context_2024} typically determines whether the \emph{freshness} or the \emph{latency} variant is more suitable.

\subsection{Conformance Checks at the Row Level}
\label{subsection:record}
This category addresses the question ``\emph{What DQ checks can be executed on the rows of a table?}''. These functionalities operate on complete rows (also termed \emph{records} or \emph{tuples}), each containing multiple values (also termed \emph{attributes}). Like value-level checks, they produce binary output (\emph{pass} or \emph{fail}) based solely on the examined row.

On top of these checks, the tools typically offer two types of table-level aggregations: computing the pass rate (i.e., number or fraction of passing rows), and identifying failing rows (all or a sample) for manual inspection by the user. Implementation-wise, these checks utilize simple SQL \texttt{SELECT} queries (or equivalents) with the negated condition in the \texttt{WHERE} clause, optionally adding \texttt{COUNT} operations for pass rate calculations.
Row checks can be summarized as:
\begin{description}
    \item[Input:] a data row (i.e., record)
    \item[Output:] pass/fail
    \item[Add-ons:] table-level aggregation (number/fraction, failing rows)
    \item[Source code:] SQL \texttt{SELECT-WHERE-COUNT} queries or equivalent
\end{description}
\autoref{table_functionalities_record} presents the low-level functionalities in this category with their implementation variants.

\begin{table*}[tb!]
    \caption{The low-level functionalities in the category of row checks with their implementation variants found in the tools examined.}
    \label{table_functionalities_record}
    \footnotesize
    \functionalitiesTablesStretch
    \begin{tabular}{ll} 
        \toprule
        \textbf{Low-level functionality}                                                                & \textbf{Implementation variants}                                                 \\
        \midrule
        \multirow[t]{1}{\functionalityColumnWidth}{\nextFunctionalityId~ row values satisfy comparison} & -                                                                                \\
        \midrule
        \multirow[t]{1}{\functionalityColumnWidth}{\nextFunctionalityId~ row satisfies SQL expression}  & \nextvariantId~ execution of \texttt{WHERE} clause before the check                                             \\
        \bottomrule
    \end{tabular}

\end{table*}

\subsubsection{Row values satisfy comparison~\ref{func:row:expression}}
\label{subsubsection:row:expression}
This low-level functionality validates relationships between values within the same row through simple comparisons. For instance, ensuring that one column value is greater than another, across all table rows.

\paragraph{Mapping to dimensions} This functionality connects to multiple dimensions: \emph{accuracy}, as violations often indicate erroneous representation of attributes of a real-world entity; \emph{consistency}, when values within a row contradict with each other; and \emph{compliance}, when business rules or policies dictate specific relationships between column values.

\subsubsection{Row satisfies SQL expression~\ref{func:row:SQL}}
\label{subsubsection:row:SQL}
This low-level functionality generalizes row-level comparisons by allowing custom SQL expressions. It provides a flexible mechanism for enforcing complex constraints and detecting errors that may not be captured by the limited expressiveness of simple comparisons between row values. Users can leverage SQL syntax to express logical conditions, mathematical operations, and string operations tailored to specific requirements.

\paragraph{Variants} The data verification scope can be restricted through the execution of \texttt{WHERE} clauses before the actual check.

\paragraph{Mapping to Dimensions} As a generalization of row-value comparisons, this functionality similarly relates to \emph{accuracy}, \emph{consistency}, and \emph{compliance}.

\subsection{Conformance Checks at the Column Level}
This category addresses the question ``\emph{What checks can run on columns of data values to assess their quality?}''. These functionalities operate at column granularity, examining a single column of a table. While most checks produce a binary outcome (\emph{pass} or \emph{fail}), few may also report the fraction of compliant values.

A key distinction from column-aggregated value checks (\autoref{subsection:value_checks}) is that checks in the current category require examining the entire column rather than individual values. For instance, validating that all column values are in increasing order inherently requires considering all values collectively, because neither single-value examination nor individual compliance is sufficient. Implementation-wise, these checks focus on computing column-wide measures or constraints, typically evaluating these results against defined thresholds. In summary, column checks are characterized by:

\begin{description}
    \item[Input:] a single column
    \item[Output:] primarily pass/fail
    \item[Source code:] computation of column measure or constraint; threshold comparison 
\end{description}
\autoref{table_functionalities_column} presents the low-level functionalities in this category along with their implementation variants as identified across the examined tools. 
\begin{table*}[tb!]
    \caption{The low-level functionalities in the category of column checks with their implementation variants found in the tools examined.~\noteAboutOrthogonal}
    \label{table_functionalities_column}
    \footnotesize
    \functionalitiesTablesStretch
    \begin{tabular}{ll} 
        \toprule
        \textbf{Low-level functionality}                                                                      & \textbf{Implementation variants}                                                         \\
        \midrule
        \multirow[t]{4}{\functionalityColumnWidth}{\nextFunctionalityId~ simple descriptive statistic checks} & \nextvariantId~ equality to a desired value                                              \\
                                                                                                              & \nextvariantId~ threshold-based comparisons                                              \\
                                                                                                              & \nextvariantId~ statistic measure computed on a sample (orthogonal)                      \\
        \midrule
        \multirow[t]{5}{\functionalityColumnWidth}{\nextFunctionalityId~ values satisfy ordering}             & \nextvariantId~ increasing/decreasing ordering                                           \\
                                                                                                              & \nextvariantId~ non-strict ordering (i.e., equality accepted between consecutive values) \\
                                                                                                              & \nextvariantId~ fixed step of increase/decrease                                          \\
                                                                                                              & \nextvariantId~ relaxed compliance to ordering (orthogonal)                              \\
                                                                                                              & \nextvariantId~ execution of \texttt{GROUP BY} clause before the check (orthogonal)      \\
        \bottomrule
    \end{tabular}
\end{table*}


\subsubsection{Simple descriptive statistic checks~\ref{func:column:statistics}}
\label{subsubsection:column:statistics}
This widely-implemented functionality performs checks on column-level statistical measures, covering a wide spectrum of descriptive statistical analysis, which is also part of data profiling. For example, in a fifty-person company's project management system, it could verify that employee assignments across projects sum to exactly fifty, assuming every employee works on exactly one project.
The examined tools support various statistics including minimum, maximum, mean, median, standard deviation, variance, and z-scores. For z-scores specifically, tools may also report the fraction of values passing the check. A comprehensive list of supported statistics can be derived from the exact names of source code elements for this functionality, presented in \autoref{appendix:function_names}.

\paragraph{Variants} 

Check outcomes are determined in several ways after computing the respective statistics. The first approach requires exact equality to a user-defined value. The second enables threshold-based comparisons (greater than. less than, between values). The generalization of these approaches allows custom callback functions that implement specific validation logic on the computed statistics. An orthogonal variant reduces the computational resources required by computing statistics on data \emph{samples}, rather than entire columns. While currently implemented only for standard deviation and variance calculations, this approach could extend to other statistics, particularly when prioritizing real-time assessment with acceptable accuracy trade-offs.

\paragraph{Mapping to dimensions} 
Descriptive statistics provide indirect insights into how accurately data represent reality. For instance, negative minimum values in a product quantity column indicate compromised \emph{accuracy}. Additionally, computing the mean value of a column requires the examination of all the values in that column. This means data values have to be coherent with other data in some context of use, in order for the check to be successful, making \emph{consistency} relevant. \emph{Compliance} applies when statistical constraints stem from business rules, such as the requirement that every employee in the fifty-person company is involved in exactly one project.

\subsubsection{Values satisfy ordering~\ref{func:column:ordering}}
\label{subsubsection:column:ordering}
This low-level functionality checks whether all consecutive values in a column satisfy a user-defined ordering, operating directly on the data values. For instance, in an IoT scenario measuring hourly household energy consumption, we expect the cumulative sum readings within a single day to be increasing.

\paragraph{Variants} The variants primarily differ in how ordering is defined and enforced. The basic ordering can be either increasing or decreasing, with additional options for strict or non-strict ordering. A more restrictive variant requires fixed increments or decrements between consecutive pairs, e.g., ensuring that the cumulative energy consumption increases by exactly 10 units hourly. Two orthogonal variants offer flexibility in application of the check. One allows for \emph{relaxed} compliance by requiring only a specified fraction of values to satisfy ordering (e.g., at least 80\% of column values must increase). The other enables pre-grouping of data through SQL \texttt{GROUP BY} clauses before the actual ordering check.

\paragraph{Mapping to dimensions} This functionality primarily connects to \emph{accuracy}, as ordering violations often indicate unrealistic data patterns, such as decreasing cumulative consumption readings. It also relates to \emph{consistency}, since values that violate the expected ordering contradict with the rest column values. \emph{Compliance} becomes relevant when ordering requirements stem from business rules or regulations.  For instance, consider a company's policy to assign increasing transaction identifiers to every new transaction generated. The identifier column is expected to have strictly increasing values, or else the business rule is violated.

\subsection{Conformance Checks at the Schema Level}
\label{subsection:schema_checks}
This category addresses the question ``What error detection checks can be executed on the schema of a database to assess its quality?''. The functionalities in this category primarily operate at the table level, i.e., investigating (all or part of) the table columns to verify that schema is correctly reflected in the data. Some variants, however, particularly those involving data type validation (cf.~\autoref{subsubsection:schema:types}), interfere with the column level as well.

\autoref{table_functionalities_schema} presents the low-level functionalities in the schema checks category with the implementation variants as identified across the examined tools.

\begin{table*}[tb!]
    \caption{The low-level functionalities in the category of schema checks with their implementation variants found in the tools examined.~\noteAboutOrthogonal}
    \label{table_functionalities_schema}
    \footnotesize
    \functionalitiesTablesStretch
    \begin{tabular}{ll} 
        \toprule
        \textbf{Low-level functionality}                                                          & \textbf{Implementation variants}                                              \\
        \midrule
        \multirow[t]{6}{\functionalityColumnWidth}{\nextFunctionalityId~ columns match schema}    & \nextvariantId~ number of columns in a dataset                                \\
                                                                                                  & \nextvariantId~ expected columns are present in the dataset                   \\
                                                                                                  & \nextvariantId~ unexpected columns are not present in the dataset             \\
                                                                                                  & \nextvariantId~ columns' indices satisfy ordering                             \\
                                                                                                  & \nextvariantId~ use of regexes to specify column names (orthogonal)           \\
                                                                                                  & \nextvariantId~ comparison against reference data (orthogonal)                \\
        \midrule
        \multirow[t]{4}{\functionalityColumnWidth}{\nextFunctionalityId~ data types match schema} & \nextvariantId~ values are of specific data type(s) and/or format(s)          \\
                                                                                                  & \nextvariantId~ distribution of data types in a column                        \\
                                                                                                  & \nextvariantId~ all table columns match schema                                \\
                                                                                                  & \nextvariantId~ data types are inferred if no schema information (orthogonal) \\
                                                                                                  & \nextvariantId~ comparison against reference data (orthogonal)                \\
        \bottomrule
    \end{tabular}

\end{table*}


\subsubsection{Columns match schema~\ref{func:schema:columns}}
\label{subsubsection:schema:columns}
This low-level functionality verifies the presence, absence, and ordering of columns in a table against an expected schema. It detects structural discrepancies, such as missing or extraneous columns, that might arise during data ingestion, integration, or preprocessing. For example, in a table used for financial reporting, it can verify that critical columns, such as the account's identifier, the transaction amount and date are present and appear in the sequence required by reporting standards.

\paragraph{Variants} The simplest variant validates just the number of columns against a desired value or threshold. Note that this is also discussed from another perspective in the low-level functionality about data element counts~\ref{func:volume:elements}~(\autoref{subsubsection:volume:elements}). A more expressive variant allows the user to define a set of the exact \emph{column names} the table at hand must have. The complementary variant uses a set of \emph{unexpected} columns that must \emph{not} be present. In a stricter flavor, the column indices are also checked, i.e., the \emph{ordering} of the columns, besides just their presence in the table. This variant is only applicable for a set of expected columns. Two orthogonal variants enhance flexibility. One enables the user to specify regular expressions for columns, rather than exact names, while the other allows for comparison against reference data.

\paragraph{Mapping to Dimensions} Since table columns represent real-world entity attributes, ensuring their presence highly relates to \emph{completeness}. \emph{Accessibility} becomes relevant with missing columns, since absent attributes are unavailable to users or downstream tasks. \emph{Compliance} applies when business rules or standards dictate column ordering or restrictions; for instance, reporting standards requiring specific column sequences or access policies prohibiting the presence of certain columns.

\subsubsection{Data types match schema~\ref{func:schema:types}}
\label{subsubsection:schema:types}
This low-level functionality ensures data type conformance with the table schema. It can prevent downstream data processing errors by ensuring that fields are correctly typed, e.g., numeric, string, or dates. While primarily operating at the table level, some variants focus on the column level as well, i.e., checking compliance to data types for single columns.

\paragraph{Variants} The simplest variant verifies type matching of (all or part of) the table columns, with tools supporting specialized formats (e.g., \texttt{strftime} or \texttt{dateutil} for dates). More details about the supported types and formats can be derived from the source code elements' names appearing in~\autoref{appendix:function_names}. Implementation typically involves attempted value parsing (conversion) in the specified format, keeping track of failing values. Note that compliance to regular expressions and patterns is also discussed from another perspective in~\ref{func:value:regex}~(\autoref{subsubsection:value:regex}). Advanced variants include analyzing the distribution of data types within columns (e.g., ensuring at least 80\% type compliance) and table-wide schema validation using JSON-like type specifications targeting all table columns. Orthogonal variants include automated type inference when schema information is missing, and comparison against reference data, where an additional option supports checking respective columns one by one via column mapping.

\paragraph{Mapping to Dimensions}
Type conformance primarily connects to \emph{accessibility}, as non-compliant values often fail to be properly displayed to humans or processed by downstream tasks; thus, they become inaccessible. The expected schema itself, though, is frequently derived from business-internal standards or communication protocols between systems (steps) in a pipeline, making \emph{compliance} also relevant. From another perspective, a mismatched data type can also indicate that the respective value cannot be corresponding to a real-world entity. For instance, non-numeric quantity values cannot represent actual quantities indicating \emph{accuracy} issues.

\subsection{Conformance Checks at the Table Level}
\label{subsection:dataset_checks}
This category addresses the question ``\emph{What data quality checks can be executed on an entire table?}''  The low-level functionalities in this category operate on complete tables (also termed \emph{datasets} in machine learning contexts). While the check outcome is primarily binary (\emph{pass} or \emph{fail}), some variants may also report the number or fraction of failing rows, or even present them to the user for manual inspection.~\autoref{table_functionalities_dataset} presents the low-level functionalities in this category along with their implementation variants as identified across the examined tools.

\begin{table*}[tb!]
    \caption{The low-level functionalities in the category of table checks with their implementation variants found in the tools examined.~\noteAboutOrthogonal}
    \label{table_functionalities_dataset}
    \footnotesize
    \functionalitiesTablesStretch
    \begin{tabular}{ll} 
        \toprule
        \textbf{Low-level functionality}                                                                     & \textbf{Implementation variants}                                               \\
        \midrule
        \multirow[t]{5}{\functionalityColumnWidth}{\nextFunctionalityId~ matching between source \& target}  & \nextvariantId~ strict (complete) matching                                     \\
                                                                                                             & \nextvariantId~ user-defined fraction of matching                              \\
                                                                                                             & \nextvariantId~ custom precision for numeric comparisons (orthogonal)          \\
        \midrule
        \multirow[t]{3}{\functionalityColumnWidth}{\nextFunctionalityId~ adjacent intervals do not overlap}  & \nextvariantId~ gaps between consecutive intervals allowed/disallowed/required \\
                                                                                                             & \nextvariantId~ zero-span interval allowed/disallowed                          \\
                                                                                                             & \nextvariantId~ execution of \texttt{PARTITION BY} clause before the check     \\

        \bottomrule
    \end{tabular}
\end{table*}


\subsubsection{Matching between tables~\ref{func:table:matching}} 
\label{subsubsection:table:matching}
This low-level functionality compares  the table at hand (termed \emph{target}) to a reference table (termed \emph{source}) at a granularity level of values (i.e., table cells). If column names differ in source and target, users can specify mappings to establish one-to-one relationships between them. This functionality computes the number or fraction of matching (or mismatching) values, which can then be evaluated against user-defined thresholds. Implementation-wise, most tools use SQL joins (or key-value equivalents) to quantify the matching.

\paragraph{Variants.} Three variants offer different matching requirements. The strictest demands \emph{complete} matching between tables. A more flexible variant allows users to specify a threshold in range $(0, 1)$, against which the computed fraction of matching values is evaluated. An orthogonal variant enables custom precision for numeric comparisons; for example, matching values up to their 4\textsuperscript{th} decimal place.

\paragraph{Mapping to Dimensions} This low-level functionality connects to multiple dimensions, depending on the context. When the source table represents ground truth (\emph{gold standard}), matching relates to \emph{accuracy}. \emph{Consistency} becomes relevant as mismatching values are not coherent with reference (source) data. The functionality also addresses \emph{completeness} by identifying values present in the source but missing from the target. From another perspective, assuming that data are expected to flow from source to target, such mismatches can indicate \emph{currentness} issues serving as (indirect) latency assessment at the target end. Additionally, missing values from the target may signal \emph{accessibility} problems, as data become unavailable to end users or downstream tasks fed with the target data.

\subsubsection{Adjacent intervals do not overlap~\ref{func:table:overlapping}}
\label{subsubsection:table:overlapping}
This low-level functionality validates that intervals defined by two columns (e.g., lower and upper bounds) do not overlap across table rows. For instance, in hotel room reservations, where each row contains check-in and check-out dates for some room, preventing overlapping bookings for the same room is crucial. In general, such overlaps could indicate logical errors, resource conflicts, or misconfigured time periods across various domains. Implementation-wise, the functionality first sorts the rows by the lower bound column, then leverages the SQL \texttt{LEAD} window function to simultaneously access adjacent rows.

\paragraph{Variants} The first variant manages gaps between adjacent intervals. The gaps can be allowed (e.g., room reservations), disallowed (e.g., shifts in a production schedule), or required (e.g., mandatory cleaning period between bookings). The second variant controls whether zero-span intervals within single rows are permitted. The third enables interval validation within distinct groups (e.g., room numbers) independently. Rows are partitioned by a specified column, and overlap validation is applied only within each group, leveraging internally an SQL \texttt{PARTITION BY} clause.

\paragraph{Mapping to Dimensions} This functionality connects to \emph{accuracy}, as overlapping intervals often indicate incorrect real-world representation. \emph{Consistency} becomes relevant because interval validation requires examining relationships between adjacent ranges, not just individual values. The functionality also relates to \emph{compliance} when domains enforce specific interval policies, such as non-overlapping booking rules.

\subsection{Distribution-based checks}
\label{subsection:distribution}
This category addresses the question ``\emph{"How can the investigation of the distribution of data be used to draw conclusions about their quality?}'' We introduce this category due to the prevalence of distribution-related functionalities across the examined tools. These checks primarily operate at the column level, i.e., investigating the distribution of single columns. The unifying characteristic is their focus on distribution analysis, rather than direct value examination. Most functionalities produce binary outcomes (\emph{pass} or \emph{fail}). Distribution checks can be characterized by:

\begin{description}
    \item[Input:] single columns
    \item[Output:] pass/fail
    \item[Source code:] distribution computation followed by threshold comparison
\end{description}
\autoref{table_functionalities_distribution} presents the low-level functionalities in this category along with their implementation variants as identified across the examined tools.

\begin{table*}[tb!]
    \caption{The low-level functionalities in the category of distribution checks with their implementation variants found in the tools examined.~\noteAboutOrthogonal}
    \label{table_functionalities_distribution}
    \footnotesize
    \functionalitiesTablesStretch
    \begin{tabular}{ll} 
        \toprule
        \textbf{Low-level functionality}                                                       & \textbf{Implementation variants}                                                      \\
        \midrule
        \multirow[t]{3}{\functionalityColumnWidth}{\nextFunctionalityId~ histogram checks}     & \nextvariantId~ only the $k$ most frequent values are considered                      \\
                                                                                               & \nextvariantId~ execution of UDF on column values before the check (orthogonal)       \\
                                                                                               & \nextvariantId~ execution of \texttt{WHERE} clause before the check (orthogonal)      \\
        \midrule
        \multirow[t]{6}{\functionalityColumnWidth}{\nextFunctionalityId~ quantiles checks}     & \nextvariantId~ user-defined single/multiple quantile(s)                              \\
                                                                                               & \nextvariantId~ equality to a desired value                                           \\
                                                                                               & \nextvariantId~ threshold-based comparison                                            \\
                                                                                               & \nextvariantId~ approximate computation of quantiles (orthogonal)                     \\
                                                                                               & \nextvariantId~ comparison against reference data (orthogonal)                        \\
        \bottomrule
    \end{tabular}
\end{table*}


\subsubsection{Histogram checks~\ref{func:distribution:histogram}}
\label{subsubsection:distribution:histogram}
This low-level functionality operates in two stages: first computing a column's distribution histogram in a single pass on the data, then enabling user queries and threshold-based checks on the resulting distribution object. Designed primarily for categorical data, it creates bins corresponding to the distinct column values. For example, consider IoT devices reporting measurements along with their operation status (e.g., ``Healthy'', ``Unhealthy'', or ``Unknown''). The functionality could verify that at any minute, at least 50\% of devices report ``Healthy'' status and fewer than 10\% report ``Unknown'', ensuring reliability for downstream analysis.

\paragraph{Variants} While the functionality is primarily targeting categorical data, one variant offers the flexibility to discretize continuous values. Discretization is achieved through user-defined functions (UDFs) which are called on every column value, applying the transformation before distribution computation. Another variant allows users to specify a parameter $k$ to consider only the top-$k$ most frequent items, ignoring less frequent values in the distribution. An orthogonal variant enables pre-filtering through SQL \texttt{WHERE} clauses, excluding specific rows from distribution computation.

\paragraph{Mapping to dimensions} Distribution analysis connects to \emph{accuracy} by providing indirect insights into how well the data represents the real-world. \emph{Consistency} becomes relevant as distribution checks require coherence of single values with the rest column values. The functionality also relates to \emph{compliance} when distribution requirements stem from business rules or key performance indicators (KPIs), as in the IoT monitoring example.

\subsubsection{Quantiles checks~\ref{func:distribution:quantiles}}
\label{subsubsection:distribution:quantiles}
This low-level functionality computes and validates user-defined quantiles of column values. For instance, it can verify that the 95\textsuperscript{th} percentile of the column values matches a desired value. 

\paragraph{Variants} The basic variants support single or multiple quantile checks, typically specified by the user in the form of \emph{percentiles}. Validation can require either exact equality to desired value or threshold-based comparisons (greater or smaller than thresholds, or within a low and a high one). 

Two orthogonal variants enhance functionality. The first enables \emph{approximate} quantile computation, improving performance at the cost of result accuracy. This approximation is implemented in two alternatives: the first leverages the \texttt{ApproximatePercentile} class of Apache Spark's \texttt{sql.catalyst} optimizer\footnote{\url{https://github.com/apache/spark/blob/master/sql/catalyst/src/main/scala/org/apache/spark/sql/catalyst/expressions/aggregate/ApproximatePercentile.scala}}, while the second leverages the KLL sketching algorithm~\cite{kll_sketch}. The second orthogonal variant enables comparison against reference data, quantifying differences between computed and reference quantiles. Users can subsequently specify maximum acceptable differences, either as absolute or percentage values.

\paragraph{Mapping to dimensions} Like distribution histogram checks, this functionality connects to \emph{accuracy}, \emph{consistency} and \emph{compliance}.

\subsection{Volume- and Cardinality-based checks}
This category addresses the question "\emph{What error detection checks on the quantity of the data serve as quality indicators?}". We introduce this category to encompass the numerous functionalities, found across the tools that assess data quality through  volume and cardinality.

While primarily operating at the column level, several checks in this category span multiple granularity levels, as shown in~\autoref{table_functionalities_to_dimensions}. The result of the check can be binary (\emph{pass} or \emph{fail}), absolute counts, or relative measures (e.g., fraction of elements meeting specific criteria).~\autoref{table_functionalities_volume_cardinalities} presents the low-level functionalities in this category along with their implementation variants as identified across the examined tools.

\begin{table*}[ht]
    \caption{The low-level functionalities in the category of column checks with their implementation variants found in the tools examined.~\noteAboutOrthogonal}
    \label{table_functionalities_volume_cardinalities}
    \scriptsize
    \functionalitiesTablesStretch
    \resizebox{\linewidth}{!}{%
        \begin{tabular}{ll} 
            \toprule
            \textbf{Low-level functionality}                                                                        & \textbf{Implementation variants}                                                              \\
            \midrule
            \multirow[t]{7}{\functionalityColumnWidth}{\nextFunctionalityId~ checks on the number of elements}      & \nextvariantId~ number of values in a column                                                  \\
                                                                                                                    & \nextvariantId~ existence of at least one value in a column                                   \\
                                                                                                                    & \nextvariantId~ number of rows in a dataset                                                   \\
                                                                                                                    & \nextvariantId~ number of columns in a table                                                  \\
                                                                                                                    & \nextvariantId~ equality to a desired value (orthogonal)                                      \\
                                                                                                                    & \nextvariantId~ threshold-based comparison (orthogonal)                                       \\
                                                                                                                    & \nextvariantId~ comparison against reference data (orthogonal)                                \\
            \midrule
            \multirow[t]{6}{\functionalityColumnWidth}{\nextFunctionalityId~ distinct elements checks}              & \nextvariantId~ number of distinct values in a column (cardinality)                           \\
                                                                                                                    & \nextvariantId~ fraction of distinct values in a column                                       \\
                                                                                                                    & \nextvariantId~ distinct values in a column cover/equal set of values                         \\
                                                                                                                    & \nextvariantId~ approximate computation of distinct values (orthogonal)                       \\
                                                                                                                    & \nextvariantId~ comparison against reference data (orthogonal)                                \\
            \midrule
            \multirow[t]{5}{\functionalityColumnWidth}{\nextFunctionalityId~ unique elements checks}                & \nextvariantId~ number of unique values in a column                                           \\
                                                                                                                    & \nextvariantId~ fraction of unique values in a column over all values (uniqueness)            \\
                                                                                                                    & \nextvariantId~ fraction of unique values in a column over distinct ones (unique value ratio) \\
                                                                                                                    & \nextvariantId~ values in a column are all unique (primary key verification)                  \\
                                                                                                                    & \nextvariantId~ number/fraction of \emph{duplicate} values in a column                        \\
                                                                                                                    & \nextvariantId~ number/fraction of duplicate rows in a table                                  \\
                                                                                                                    & \nextvariantId~ number of duplicate columns in a table                                        \\
                                                                                                                    & \nextvariantId~ comparison against reference data (orthogonal)                                \\
            \midrule
            \multirow[t]{4}{\functionalityColumnWidth}{\nextFunctionalityId~ most common elements checks}           & \nextvariantId~ column values are not identical                                               \\
                                                                                                                    & \nextvariantId~ fraction of most common value over all values                                 \\
                                                                                                                    & \nextvariantId~ number of columns in a table with constant values                             \\
                                                                                                                    & \nextvariantId~ execution of \texttt{GROUP BY} clause before the check (orthogonal)           \\
                                                                                                                    & \nextvariantId~ comparison against reference data (orthogonal)                                \\
            \midrule
            \multirow[t]{12}{\functionalityColumnWidth}{\nextFunctionalityId~ missing elements checks}              & \nextvariantId~ a column has no missing values                                                \\
                                                                                                                    & \nextvariantId~ a column has only \texttt{NULL} values                                        \\
                                                                                                                    & \nextvariantId~ number/fraction of missing values in a column                                 \\
                                                                                                                    & \nextvariantId~ number/fraction of rows with missing values in a dataset                      \\
                                                                                                                    & \nextvariantId~ number/fraction of empty rows in a rows                                       \\
                                                                                                                    & \nextvariantId~ number/fraction of rows with missing values in all/any columns of a set       \\
                                                                                                                    & \nextvariantId~ number/fraction of missing values in a table                                  \\
                                                                                                                    & \nextvariantId~ number of columns with missing values (empty or at least one)                 \\
                                                                                                                    & \nextvariantId~ calculation of \emph{non-missing} values, instead of missing (orthogonal)     \\
                                                                                                                    & \nextvariantId~ custom-defined representation(s) of missing values (orthogonal)               \\
                                                                                                                    & \nextvariantId~ execution of \texttt{WHERE} clause before the check (orthogonal)              \\
                                                                                                                    & \nextvariantId~ execution of \texttt{GROUP BY} clause before the check (orthogonal)           \\
                                                                                                                    & \nextvariantId~ comparison against reference data (orthogonal)                                \\
            \midrule
            \multirow[t]{1}{\functionalityColumnWidth}{\nextFunctionalityId~ checks on representations for missing} & \nextvariantId~ custom-defined representation(s) of missing values (orthogonal)               \\
                                                                                                                    & \nextvariantId~ comparison against reference data (orthogonal)                                \\
            \bottomrule
        \end{tabular}
    }
\end{table*}


\subsubsection{Checks on the number of elements~\ref{func:volume:elements}}
\label{subsubsection:volume:elements}
This low-level functionality quantifies data elements at various granularity levels (\emph{rows}, \emph{columns}, or \emph{table}), enabling comparison against expected counts or threshold-based constraints.

\paragraph{Variants} Variants of this low-level functionality offer different perspectives in evaluating the quantification of data elements across granularity levels:
\begin{itemize}
    \item \textit{column} level: number of values in a column or verification that at least one value exists in a column. The latter ensures bare minimum availability of the respective attribute; 
    \item \textit{row} level: number of columns (i.e., attributes) per row;
    \item \textit{table} level: number of columns in the table.
\end{itemize}
An orthogonal variant enables reference data comparison, where constraints apply to the difference between actual and reference quantities, rather than direct value counts.

\paragraph{Mapping to Dimensions} Volume-based checks primarily connect to \emph{completeness}, providing insights into whether data adequately represent all expected real-world attributes or entities. They also relate to \emph{accessibility}, particularly its availability aspect, as absent elements are inherently inaccessible to users or downstream tasks.

\subsubsection{Distinct elements checks~\ref{func:volume:distinct}}
\label{subsubsection:volume:distinct}
This low-level functionality computes the number of \emph{distinct} data elements, primarily operating at the column level, i.e., distinct values in a column. Tools frequently term this functionality as (column) \emph{cardinality}. Here, \emph{distinct} refers to elements appearing at least once (cf. \emph{unique} elements in~\autoref{subsubsection:volume:unique} has a slightly different meaning). For example, in sequence $[a, a, b]$ the distinct values are $a$ and $b$. The computed cardinality can then be validated against user-defined counts or thresholds.

\paragraph{Variants} Beyond basic distinct value counts, the \emph{fraction} of distinct values relative to total column values can be computed. A stricter variant validates set relationships, ensuring distinct values either cover or exactly match a specified value set (improper or proper subset, respectively). One orthogonal variant employs the HyperLogLog++ sketching algorithm~\cite{hll_plus_plus_2013} for approximate computation, trading accuracy for reduced computational cost. The other enables comparison against reference data, validating cardinality against ideal or desired states.

\paragraph{Mapping to Dimensions}
Verifying the number of distinct values adds more specific requirements on top of the basic volume checks towards data \emph{completeness}; not only has the volume of data be the expected one, but also their cardinality needs to meet constraints. The functionality also connects to \emph{accessibility} as cardinality violations indicate non-represented (thus inaccessible) real-world entities. It also relates to \emph{consistency}, as cardinality checks require coherence across all values; individual values, though correct in isolation, must collectively satisfy cardinality requirements.

\subsubsection{Unique elements checks~\ref{func:volume:unique}}
\label{subsubsection:volume:unique}
This functionality quantifies \emph{unique} data elements, i.e., those appearing \emph{exactly} once. For instance in sequence $[a, a, b]$, only $b$ is unique. This concept is directly linked to duplicate detection, redundant information, and validating primary key constraints in relational tables.

\paragraph{Variants}
The functionality variants operate at multiple granularity levels:
\begin{itemize}
    \item \textit{column} level: The simplest variant verifies constraints on the number of unique values in a column. Other variants compute the fraction of unique values over all column values $\frac{|unique|}{n_{col}}$ (termed \emph{uniqueness}) or over only distinct ones $\frac{|unique|}{|distinct|}$ (termed \emph{unique value ratio}). Primary key verification can be considered a stricter version of these variants, asking for \emph{all} column values to be unique (i.e., $uniqueness = unique~value~ratio = 1$). Another variant implements checks on the number or fraction of \emph{duplicate} values in the column.
    \item \textit{table} level: The number of duplicate rows is computed, or their fraction over the total number of rows in the table. Another variant computes the number of duplicate \emph{columns} in the table, examining all-to-all pairs.
\end{itemize}
All these variants can be combined with comparisons against reference data.

\paragraph{Mapping to Dimensions}
\emph{Completeness}, \emph{consistency} and \emph{accessibility} are the connected dimensions in descending order of connection strength, similarly to distinct value checks. Consistency, though, has a heightened relevancy, especially with the primary key verification variant. In some cases, \emph{accuracy} can also become relevant, particularly when unexpected duplicate values indicate erroneous representation of the real world; for example, when a distinct entity of the real-world is mistakenly mapped to an existing key of another entity.

\subsubsection{Most common elements checks~\ref{func:volume:common}}
\label{subsubsection:volume:common}
This functionality focuses on the frequency of most common data elements, primarily at the column level. It helps detect dominant categories and assess data uniformity. For example, in a customer feedback analysis, it can ensure that the column representing feedback type has not an overwhelming majority of ``Positive'' responses, which could indicate sampling bias.

\paragraph{Variants} This functionality offers several validation approaches. The basic variant ensures that column values are not identical. The generic form of this variant is verifying constraints on the fraction of the most common column value over the total number of column values. The difference between them is that the former fails if the fraction result is other than 1 (i.e., the frequency of the most common value equals the number of values in total), while in the latter, the user can define a custom threshold (e.g., the check fails if fraction $>0.8$). At the table level, another variant counts columns containing constant values, comparing it with a user-defined threshold.

Two orthogonal variants enhance functionality. One enables pre-processing through an SQL \texttt{GROUP BY} clause, which groups rows and potentially produces aggregated fields. The other supports comparison against reference data, enabling user to specify thresholds on the difference.

\paragraph{Mapping to Dimensions}
This functionality primarily connects to \emph{completeness}, as the frequency of most common data elements can directly reflect the degree to which real-world entities or attributes are represented in the data. \emph{Consistency} becomes relevant because frequency patterns require coherence across values. The functionality can also relate to \emph{accuracy}, particularly when unusual frequency patterns suggest data issues. For instance, in healthcare records, placeholder values such as ``999'' might be used to indicate missing or unknown blood pressure readings. If this placeholder appears with an unusually high frequency, it could point to problematic data entry or equipment malfunctions during data collection, compromising the accurate representation of patient health states.

\subsubsection{Missing elements checks~\ref{func:volume:missing}}
\label{subsubsection:volume:missing}
This low-level functionality quantifies and checks the number of missing data elements across multiple granularity levels. 

\paragraph{Variants}
The following variants were found across the examined tools:
\begin{itemize}
    \item \textit{column level}: Some implemented checks offer support for verifying that a column has no missing values or is completely empty. The generic form these variants can take is verifying constraints on the number or fraction of missing values in the column. 
    \item \textit{row level}: The first variant calculates the number or fraction of rows with at least one missing value in any column (attribute). The second variant differentiates in searching for \emph{completely} empty rows, i.e., records with all attributes missing. A compromise approach in-between them enables users to define a set of \emph{important} columns, i.e., columns that must not have missing values. The two alternatives for the latter check relate to whether it fails when missing values occur in either \emph{all} or \emph{any} of the important columns.
    \item \textit{table level}: The first variant measures the number of detects values in the table overall, regardless of rows and columns. In this case, the table is treated as a collection of individual data cells. The second variant focuses on the number of \emph{columns} with missing values, with the two extreme cases to be empty columns on the one hand, and columns with at least one missing value on the other.
\end{itemize}

Several orthogonal variants enhance functionality. The first inverts the measure of missing data, counting \emph{non-missing} values. The second variant offers flexibility in terms of what is treated as missing. While some tools support only universally accepted representations for missing values, such as ``NULL'', ``NaN'', etc., other tools enable users to define custom representations, such as empty strings, ``-'', or any case-specific ones. Another two orthogonal variants enable data preprocessing through SQL \texttt{WHERE} and \texttt{GROUP BY} clauses, respectively, before the actual check. Comparing missing data elements against reference data remains an option, as well.

\paragraph{Mapping to Dimensions} This functionality primarily connects to \emph{completeness} through direct missing value quantification. It also relates to \emph{accuracy}, as missing values denote absent or unknown real-world attributes that fail to be truthfully represented. \emph{Accessibility} also becomes relevant since missing values are unavailable to users or downstream tasks.

\subsubsection{Checks on representations for missing~\ref{func:volume:representation_missing}}
\label{subsubsection:volume:representation_missing}
This functionality counts the number of distinct representations used to denote missing values in a table and compares it to user-defined desired values or thresholds. This helps to identify inconsistent practices about how missing or undefined data are handled.

\paragraph{Variants} The first variant allows users to define custom placeholders for missing values that are also detected besides ``NULL'' or ``NaN'' values. The second variant compares the number of representations found in the data at hand to the respective measurement in a reference table. Both variants are orthogonal.

\paragraph{Mapping to Dimensions} Ensuring uniform representation(s) for missing information is clearly a step towards improved data \emph{consistency}. Data values that represent absent information have to be coherent with the rest of the data at hand.

\subsection{Correlation-based checks}
\label{subsection:inter_column}
This category addresses the question ``\emph{What are error detection checks about relationships between columns of a table?}''. These functionalities primarily operate at the column level, examining pairs of columns, with few variants extending to the table level, as shown in~\autoref{table_functionalities_to_dimensions}. Typically, the input to these checks is two columns of the same table and the output is binary, \emph{pass} or \emph{fail}.~\autoref{table_functionalities_inter_column} presents the low-level functionalities in this category along with their implementation variants.

\begin{table*}[tb!]
    \caption{The low-level functionalities in the category of inter-column checks with their implementation variants found in the tools examined.~\noteAboutOrthogonal}
    \label{table_functionalities_inter_column}
    \footnotesize
    \functionalitiesTablesStretch
    \begin{tabular}{ll} 
        \toprule
        \textbf{Low-level functionality}                                                            & \textbf{Implementation variants}                                                   \\
        \midrule
        \multirow[t]{8}{\functionalityColumnWidth}{\nextFunctionalityId~ correlation checks}        & \nextvariantId~ Pearson correlation                                                \\
                                                                                                    & \nextvariantId~ Spearman's rank correlation                                        \\
                                                                                                    & \nextvariantId~ Kendall's $\tau$                                                   \\
                                                                                                    & \nextvariantId~ Cram\'er's $V$                                                     \\
                                                                                                    & \nextvariantId~ automated discovery of highly correlated columns (orthogonal)      \\
                                                                                                    & \nextvariantId~ correlation between target \& prediction (ML-specific, orthogonal) \\
                                                                                                    & \nextvariantId~ correlation between target \& features (ML-specific, orthogonal)   \\
                                                                                                    & \nextvariantId~ comparison against reference data (orthogonal)                     \\
        \midrule
        \multirow[t]{1}{\functionalityColumnWidth}{\nextFunctionalityId~ mutual information checks} & -                                                                                  \\
        \bottomrule
    \end{tabular}
\end{table*}

\subsubsection{Correlation~\ref{func:correlation:correlation} \& mutual information~\ref{func:correlation:mutual} checks}
\label{subsubsection:correlation:correlation} We present together the low-level functionalities about correlation and mutual information since there is a significant degree of similarity between them. Their difference lies in the computed measure of relationship between the input columns (correlation and mutual information respectively). Both functionalities operate on derived measures rather than raw values, validating these measures against user-defined constraints.

\paragraph{Variants} We present here the variants for the correlation functionality, since no variants were found for the mutual information one. These variants primarily feature different correlation alternatives, computed on pairs of the input columns. \emph{Pearson}, \emph{Spearman's} (rank), \emph{Kendall's} (rank) and \emph{Cram\'er's} correlation alternatives were found across the examined tools. Users can specify either exact target values or thresholds for these measures.

Four orthogonal variants extend functionality. The first offers automated discovery of highly correlated columns, accepting a whole table as input (table-level check). Two other variants offer functionality tailored to machine learning scenarios through calculating the correlation between the target column and either the prediction, or the features respectively. We note that the correlation between target and prediction can provide insights about the performance of a machine learning model, while the correlation between target and features can reveal columns (attributes) strongly coupled with the ground truth the model is expected to predict. The last variant supports comparisons against reference data, enabling the user to specify thresholds on the computed difference.

\paragraph{Mapping to dimensions} These functionalities connect to \emph{accuracy}, as inter-column relationships often reflect real-world associations. They also relate to \emph{consistency}, as failed checks indicate contradictions between column values.

\subsection{ML-Oriented Checks}
\label{subsection:ml-oriented}

The low-level functionalities in this category share their close relation to machine learning (ML) downstream tasks. They can provide insights into the question ``What error detection checks can be executed on a dataset, before it is fed to some ML downstream application to ensure its integrity?''. While primarily operating at the table granularity level, some variants, especially related to data drift detection, can also operate on standalone columns (column-level).~\autoref{table_functionalities_ml} lists the low-level functionalities in this category along with their implementation variants.

\begin{table*}[tb!]
    \caption{The low-level functionalities in the category of ML-oriented checks with their implementation variants found in the tools examined.~\noteAboutOrthogonal}
    \label{table_functionalities_ml}
    \footnotesize
    \functionalitiesTablesStretch
    \begin{tabular}{ll} 
        \toprule
        \textbf{Low-level functionality}                                                          & \textbf{Implementation variants}                                                      \\
        \midrule
        \multirow[t]{1}{\functionalityColumnWidth}{\nextFunctionalityId~ anomaly detection}       & -                                                                                     \\
        \midrule
        \multirow[t]{1}{\functionalityColumnWidth}{\nextFunctionalityId~ targets do not conflict} & -                                                                                     \\
        \midrule
        \multirow[t]{6}{\functionalityColumnWidth}{\nextFunctionalityId~ data drift detection}    & \nextvariantId~ ad-hoc ($\mu \pm n \sigma$)                                           \\
                                                                                                  & \nextvariantId~ statistical testing (Kolmogorov-Smirnov, Chi-squared)                 \\
                                                                                                  & \nextvariantId~ distance between distributions (PSI, SWD, KL diverg., Jensen–Shannon) \\
                                                                                                  & \nextvariantId~ drift detection on embeddings data (classification, distance, MMD)    \\
                                                                                                  & \nextvariantId~ automated selection of method, based on data type, size, cardinality  \\
                                                                                                  & \nextvariantId~ number/fraction of columns with data drift                            \\
        \bottomrule
    \end{tabular}
\end{table*}


\subsubsection{Anomaly Detection~\ref{func:ml_oriented:anomaly}} 
\label{subsubsection:ml_oriented:anomaly}
This low-level functionality detects anomalies in the table at hand, by comparing either the raw data values or measures computed on them against user-defined thresholds or historical reference data. Only one tool was found to offer anomaly detection functionality, while several other tools claim to incorporate it as an advanced feature in their premium version, which is subject to paid subscription\footnote{We were not able to verify the exact anomaly detection functionality offered in these cases, since the scope of our survey was examining the source code of the investigated functionalities.}. The recorded anomaly detection functionality is based on a wide range of built-in (heuristic, non-ML) anomaly detection strategies, covering both absolute and relative comparisons, and offering support for batch (primarily) and streaming data.   

\paragraph{Mapping to Dimensions} This functionality connects to \emph{accuracy}, since anomalies are ultimately erroneous or unexpected values that fail to truthfully represent the intended real-world attribute. It also relates to \emph{consistency}, because an anomaly is apparently incoherent and contradicting with the rest of the data at hand.

\subsubsection{No Conflicts in Target Label/Value~\ref{func:ml_oriented:conflict}} 
\label{subsubsection:ml_oriented:conflict}
This low-level functionality ensures the absence of conflicts within the target labels (classification) or values (regression) in a table (dataset) used for model training. Here, a \emph{conflict} is defined as the existence of different labels or target values for the same set of attribute values, i.e., rows with exactly the same column values except for the one defined as target by the user. 

\paragraph{Mapping to Dimensions} 
This functionality primarily connects to \emph{consistency}, since conflicting rows are not coherent with the rest of the data at hand. \emph{Accuracy} becomes also relevant, particularly when rows represent real-world entities, since conflicts indicate untruthful capturing of real-world attributes in at least one of the involved rows.

\subsubsection{Data drift detection~\ref{func:ml_oriented:drift}}
\label{subsubsection:ml_oriented:drift}
This low-level functionality aims to detect data drift (i.e., significant change in data distribution) on the data at hand and requires the existence of reference data; it can also be seen as an anomaly detection functionality. The latter are considered as the ideal or desired state the data at hand are expected to be. For instance, this low-level functionality could be used in a predictive maintenance (PdM) industrial scenario, where the utter goal is to detect failures before they actually occur. Suppose that several sensors are spread across the multiple components of a production line, taking measures periodically about temperature, vibration, etc. on the respective component. These measurements are gathered and analyzed centrally in order to detect anomalies that can lead to failures. In PdM scenarios, it is common that some change (e.g., repair) in one component of the production line causes significant changes in the measurements taken at neighboring components that have not undergone any change~\cite{zenisek_pdm_concept_drift_2019}. Provided that some reference data about standard operation of each component are available, this low-level functionality could be employed to detect that untouched components have measurements that are probably from another distribution than the normal, expected one; raising timely alerts for manual inspection or system restart, etc.

\paragraph{Variants} Several variants regarding the way data drift is detected were found in the source code of the examined tools. The first variant detects data drift \emph{ad-hoc} by computing the number of standard deviations the current mean value differs from the mean value of the reference data. In particular, data drift is detected iff \(
    \mu > \mu_{ref} + n\sigma_{ref} \;\;\;\text{or}\;\;\; \mu < \mu_{ref} - n\sigma_{ref},
\)
where $\mu$ is the mean value of the data at hand, $\mu_{ref}$ is the mean value of the reference data, $\sigma_{ref}$ is the standard deviation of the reference data and $n$ is a user-defined positive integer.

The second variant uses statistical testing to define how likely is that the data at hand and the reference data are drawn from the same distribution. Kolmogorov-Smirnov (KS) and Chi-squared tests are both found in terms of implementation, letting the user define an upper threshold on the test result. The third variant uses distribution distance measures to quantify the difference between the distribution of the data at hand and the distribution of the reference data. Population Stability Index (PSI), Sliced Wasserstein Distance (SWD), Kullback–Leibler (KL) divergence and Jensen-Shannon divergence are the distribution distance measures employed by the tools implementing this alternative.

The fourth variant is tailored to embeddings, i.e., vectors representing real objects, such as text, images, etc., and are designed to be consumed by machine learning applications. The following three methods were found for detecting data drift in embeddings data:
\begin{enumerate}
    \item \emph{Binary classification}: a linear binary classifier is trained from scratch to distinguish between the data at hand and the reference data. The area under curve (AUC) of the classifier's ROC curve is considered as the ``data drift score'', enabling the user to define thresholds on this value.
    \item \emph{Distance of mean embeddings}: the mean embedding is calculated on the data at hand and the reference data and the distance between them is computed. The available distance/similarity metrics between the two vectors are \emph{euclidean}, \emph{manhattan}, \emph{chebysev} distances and \emph{cosine} similarity. The user can define thresholds on the distance result.
    \item \emph{Maximum Mean Discrepancy (MMD)}: MMD is computed between the data at hand and the reference data. The user can define thresholds on the result.
\end{enumerate}

Orthogonal variants of this low-level functionality include automated selection of the data drift detection method, based on the type (e.g., categorical), size and cardinality of the data at hand. This hides from the user the complexity of defining the data drift detection method and the varying parameters required for each method to properly execute the check. In this case, the tool developers have a predefined set of default parameters and execution falls back to them, unless otherwise specified by the user. Lastly, an orthogonal variant that is more relevant to the \emph{table} granularity level, rather than the column one, is computing the number or fraction of columns for which data drift is detected. In this variant, a reference \emph{table} is required and is compared against the \emph{table} at hand. There is also the option for the user to define a mapping between the columns of the table at hand and the reference table. After the mapping establishment, the tool examines for data drift all the respective column pairs among the tables and reports the number or fraction of columns with data drift detected. We consider this variant as orthogonal, since the drift detection method can be any of the aforementioned variants.

\paragraph{Mapping to dimensions} \emph{Accuracy}, \emph{consistency} and \emph{compliance} are closely related to distribution-based checks, as explained in~\autoref{subsection:distribution}.

\section{Discussion: materialization of DQ dimensions}
\label{section:materialization_dimensions}

In this section, we reflect on the findings of our survey by starting from the six selected DQ dimensions towards the identified low-level functionalities to shed light on the materialization of DQ dimensions in practice in terms of the checks involved. Recall that the reason we focus on these dimensions is that they correspond to the inherent DQ data properties and are targeted in an application-independent manner by the examined tools.  For each selected dimensions, we connect the dots between their theoretical definitions and the low-level functionalities that contribute to their materialization. As presented in the previous section, several functionalities are shared across multiple dimensions. Across these functionalities, we identified two different approaches:
\begin{itemize}
    \item \emph{Direct assessment}: checking actual data values against specific criteria, e.g., validating an email format to assess accuracy, or checking for missing values to assess completeness;
    \item \emph{Indirect assessment}: checking metadata or patterns derived from the data (i.e., data profiling results), e.g., examining cardinalities to assess completeness.
\end{itemize}
Orthogonally, for several checks, a reference dataset representing the ground truth is employed. In alignment to the categorization from \autoref{table_functionalities_to_dimensions}, the materialization can progress through increasing granularity levels: from \emph{value level}, over \emph{row} and \emph{column} level, to \emph{table} level assessment. 
Starting from single values and gradually expanding the validation scope towards the whole table can add contextual information crucial for  uncovering DQ issues that went unnoticed at the previous level.

\subsection{Accuracy}
\label{subsection:materialization:accuracy}
Data accuracy is defined as ``\emph{the degree to which data has attributes that correctly represent the true value of the intended attribute of a concept or event in a specific context of use. It has two main aspects, syntactic and semantic accuracy}''~\cite{iso25012}. Note that other definitions for accuracy also exist in literature (e.g.,~\cite{Haegemans_2016}), but the essence remains the same.

\paragraph{Direct assessment} At the value level, direct accuracy assessment can be achieved through functionalities that examine either the syntax or semantics of individual values. Regular expressions, formats/patterns~\ref{func:value:regex}~\ref{func:value:string_length}, and data types~\ref{func:schema:types} are the most common ways to check the syntax, while ranges or sets of accepted values~\ref{func:value:range_set} are most common to check semantics. Although many DQ problems occur at the value level (e.g., wrong data entry), contextual information from the \emph{row level} ~\ref{func:row:expression}~\ref{func:row:SQL} (e.g., match of city and zip code) or \emph{column level}~\ref{func:column:ordering} (e.g., comparison of distribution) is often required for successful detection.
The spectrum of direct accuracy assessment also entails \emph{table level} validations~\ref{func:table:matching}~\ref{func:table:overlapping}~\ref{func:ml_oriented:conflict}, which can provide even semantically richer insights on data accuracy. Note that the existence of a reference table, treated as the source of truth, may be required~\ref{func:table:matching}.

\paragraph{Indirect assessment} Here, the most common checks are on \emph{single-column} data profiling results, including statistics~\ref{func:column:statistics}, distribution~\ref{func:distribution:histogram}~\ref{func:distribution:quantiles}, and data drift detection~\ref{func:ml_oriented:drift}, as well as cardinalities~\ref{func:volume:unique}~\ref{func:volume:common}~\ref{func:volume:missing}. Inter-column relations~\ref{func:correlation:correlation}~\ref{func:correlation:mutual} can also indirectly point to accuracy issues, extending the scope of investigation in-between the column and table level. Anomaly detection~\ref{func:ml_oriented:anomaly} completes the spectrum of accuracy assessment from the table perspective.

\subsection{Completeness}
\label{subsection:materialization:completeness}
Data completeness is defined as ``\emph{the degree to which subject data associated with an entity has values for all expected attributes and related entity instances in a specific context of use}''~\cite{iso25012}.

\paragraph{Direct assessment} Quantification and verification of data volume~\ref{func:volume:elements} and dimensionality~\ref{func:schema:columns}, or identification of missing values~\ref{func:volume:missing} can directly assess completeness. Note that missing values can sometimes be represented by placeholders~\ref{func:volume:common} such as ``NaN'', ``0000'', or ``January 1st, 1970''. In scenarios where a ground truth data source is available, e.g., similar to those typically employed in data integration and entity resolution pipelines to reconcile differences in the representation of the same entity, completeness assessment at the \emph{table} level can rely on comparative results~\ref{func:table:matching}. This is particularly useful in data integration and transformation workflows, ensuring no data loss or omissions during the process.

\paragraph{Indirect assessment} Our findings reveal a family of functionalities that indirectly target completeness assessment by operating on data profiling results, such as distinct ~\ref{func:volume:distinct} or unique \& duplicate~\ref{func:volume:unique} values. These checks aim to add more depth and capture subtler gaps in the data at hand. For instance, although missing values (part of direct assessment) clearly indicate non-present information (i.e., an apparent gap in the data), quantifying the unique elements can capture errors even when the overall data volume matches the expected, due to duplicates (i.e., subtler gaps).

\subsection{Consistency}
\label{subsection:materialization:consistency}
Data consistency is defined as ``\emph{the degree to which data has attributes that are free from contradiction and are coherent with other data in a specific context of use. It can be either or both among data regarding one entity and across similar data for comparable entities}''~\cite{iso25012}. While consistency can be assessed (in accordance with the other dimensions) at different granularity levels, our findings indicate an increased importance of the assessment at the \emph{column} level.

\paragraph{Direct assessment} Consistency of single values can be assessed through ensuring referential integrity~\ref{func:value:range_set}, where a set of accepted references is required. Simple~\ref{func:row:expression} or more complex~\ref{func:row:SQL} relationships between row values directly target consistency at the \emph{row level}, while verifying the ordering of column values~\ref{func:column:ordering} extends to the \emph{column level}. At the \emph{table level}, consistency can be translated to non-overlapping intervals between adjacent rows~\ref{func:table:overlapping}, capturing logical or temporal continuity. Complex logic for managing gaps between intervals can be tailored to fit various use cases. Additionally, particularly for machine learning applications, consistency may take the form of non-contradicting training examples~\ref{func:ml_oriented:conflict}.

\paragraph{Indirect assessment} Descriptive statistics~\ref{func:column:statistics}, distribution insights~\ref{func:distribution:histogram}~\ref{func:distribution:quantiles}~\ref{func:ml_oriented:drift} and cardinality-related data profiling~\ref{func:volume:distinct}~\ref{func:volume:unique}~\ref{func:volume:common} can serve as indirect consistency indicators. Note that especially quantifying unique data elements~\ref{func:volume:unique} can be crucial for the consistency of primary keys. Standardized handling of missing values~\ref{func:volume:representation_missing} can partially contribute to achieving consistency, as well. Inter-column relationships~\ref{func:correlation:correlation}~\ref{func:correlation:mutual} extend this assessment to multiple columns, offering a richer context for understanding coherence.

\subsection{Currentness}
\label{subsection:materialization:currentness}

Data currentness is defined as ``\emph{the degree to which data has attributes that are of the right age in a specific context of use}''~\cite{iso25012} and typically requires a reference timestamp against which the data is compared to.

\paragraph{Direct assessment} Checking if data values align with expected timeframes~\ref{func:value:timestamp} squarely assesses currentness. The recorded variants indicate that subtle modifications to the reference timestamp, against which data are compared, can tailor the assessment towards data \emph{freshness} or \emph{latency}.

\paragraph{Indirect assessment} Especially for latency, the degree of matching between two tables~\ref{func:table:matching} can provide insights from a workflow perspective, taking into account contexts where data are expected to flow seamlessly from one source to another. 

\subsection{Accessibility}
\label{subsection:materialization:accessibility}

Data accessibility is defined as \emph{``the degree to which data can be accessed in a specific context of use, particularly by people who need supporting technology or special configuration because of some disability''}~\cite{iso25012}. In our analysis, we primarily focus on the first part of the definition, since it can inherently be quantified in an application-agnostic manner. Unfortunately, although of heightened importance, accessibility in the context of people with disabilities is not as straightforward. Thus, here, accessibility relates to whether data can be retrieved, interpreted, and used effectively by users or downstream tasks. Our findings reveal that accessibility can be assessed with functionalities that check whether data is physically present and also properly structured. Note that while the first aspect has a high overlap with the completeness dimension, the second one relates to compliance.

\paragraph{Direct assessment} Checking conformance to data types and specific formats~\ref{func:schema:types} can directly point to potential accessibility issues from the \emph{structural} perspective. This can be of great benefit to downstream systems expecting to ingest the data in a specific format. Other schema information, such as the presence, absence and/or order of columns~\ref{func:schema:columns}, also contribute to the assessment of accessibility. However, non-present data are also inaccessible after all. Checking the quantity of the present data elements~\ref{func:volume:elements} or the missing ones~\ref{func:volume:missing} can provide insights about gaps that impede data retrieval from this perspective. At the table level, accessibility can be assessed through the successful matching between tables~\ref{func:table:matching}; useful for ensuring that data tables preserve their structure and are compatible with downstream tasks during transformations, or integrations from various sources.

\paragraph{Indirect assessment} Quantifying the distinct~\ref{func:volume:distinct} and unique~\ref{func:volume:unique} data elements can also shed light on accessibility (primarily the non-present information aspect) as part of a more in-depth, indirect approach. 

\subsection{Compliance}
\label{subsection:materialization:compliance}
Data compliance is defined as ``\emph{the degree to which data has attributes that adhere to standards, conventions or regulations in force and similar rules relating to data quality in a specific context of use}''\cite{iso25012} and is therefore inherently rule-driven.

\paragraph{Direct assessment} Adherence to specific formats~\ref{func:value:string_length}, patterns~\ref{func:value:regex}, predefined data types~\ref{func:schema:types}, or accepted ranges/sets~\ref{func:value:range_set} directly assess compliance, ensuring the alignment of individual data \emph{values} with expected regulatory or domain constraints. At the \emph{row} level, compliance is assessed through logical expressions~\ref{func:row:expression}~\ref{func:row:SQL} such as ranges, dependencies, or calculated fields that must meet predefined criteria. At the \emph{column} level, it can take the form of verifying ordering of values~\ref{func:column:ordering}. Verifying relationships between rows, such as checking non-overlapping intervals~\ref{func:table:overlapping} extend to the \emph{table level}, along with functionalities that validate the presence of mandatory columns~\ref{func:schema:columns}, ensuring that a table meets schema requirements dictated by external standards or agreements.

\paragraph{Indirect assessment} Examination of descriptive statistics~\ref{func:column:statistics}, distribution~\ref{func:distribution:histogram}~\ref{func:distribution:quantiles}~\ref{func:ml_oriented:drift}, cardinality~\ref{func:volume:distinct}, and uniqueness~\ref{func:volume:unique} can indirectly point to compliance violations, primarily at the \emph{column level}. 

\subsection{Synthesis of Findings}

Based on our findings, we draw the following conclusions (C):

\begin{description}[leftmargin=17pt]
    \item[C1:] \textbf{Fragmented data verification landscape}. Each examined DQ tool implements a diverse subset of low-level functionalities, with significant overlap between them, but no tool covers the entire  spectrum of functionalities observed. Attaining a comprehensive understanding of data verification capabilities available in current tools can be time- and effort-consuming, especially for practitioners selecting the right tool for a given use case. 
    \item[C2:] \textbf{Non-standardized terminology}. Similar functionalities at the source code level are found under diverse names across tools (e.g., \emph{recency} and \emph{freshness} for~\ref{func:value:timestamp}). Conversely, similar terms do not necessarily correspond to the same low-level functionality (e.g., \emph{invalid} as part of value-level regex matching~\ref{func:value:regex} and \emph{validation} as part of row-wise SQL expressions~\ref{func:row:SQL}). Transferring experiences and understanding can therefore be very challenging between practitioners. For our survey, detailed source code examination was required to manually align the diverse terms to the same functionalities.
    \item[C3:] \textbf{Perspective-dependent mappings}. Since each DQ dimension can be viewed from many perspectives (e.g., completeness in terms of missing values or population completeness), connecting the dots to the low-level functionalities is heavily dependent on the respective perspective from which each functionality is considered.
    This leads to a many-to-many relationship, which however provides valuable insights for the refinement of scientific DQ assessment models.
    \item[C4:] \textbf{Relationships between dimensions}. Reflecting on the way DQ dimensions are materialized in current DQ tools, some dimensions are inter-connected in a non-intuitive way. For example, completeness and accessibility share a close relation to physically non-present information, while accessibility and compliance inherently connect to conformance of values to formats or rules. These findings contribute to the further research on DQ dimensions and potentially trigger a critical reflection on the general idea to ``start a DQ program by selecting DQ dimensions''. 
\end{description}

\subsubsection{How to use this survey}
\autoref{table_functionalities_to_dimensions} serves as starting point and overview of this survey. It lists all the recorded low-level functionalities, introduces the many-to-many mapping and provides links for easy navigation through the content. However, we strongly encourage readers to study the thorough discussion of the low-level functionalities in~\autoref{section:survey_findings}, before moving to the wrap-up discussion in~\autoref{section:materialization_dimensions}, since the implementation variants and the explained mapping to dimensions are meant to pave the way for a deeper understanding of the conclusions. This strongly applies to data practitioners, as well. 
We recommend to first look at~\autoref{table_functionalities_to_dimensions} to identify potentially useful functionalities for a specific use case and, second, to explore the detailed discussion for these functionalities, before implementing any of the checks on the data.
The questions at the beginning of each subsection in~\autoref{section:survey_findings} aim to support practitioners in identifying whether the respective content aligns with their needs or can be skipped.

\section{Related work}
\label{section:related_work}
In this section, we first compare our work to other surveys that investigate DQ tools (cf. \autoref{section:related_work_DQ_surveys}) and second, provide an overview on relevant research aiming to define DQ dimensions and putting them in a practical context (cf. \autoref{section:related_work_DQ_dimensions}).

\subsection{Related surveys of DQ tool functionalities}
\label{section:related_work_DQ_surveys}
In 2005, \citet{Barateiro_DQSurvey_2005} compared the technical functionalities of 9 academic and 28 commercial DQ tools. \citet{Pushkarev_DQSurvey_2010} evaluated 7 open-source or freely available DQ tools in terms of \textit{performance criteria and \textit{core functionalities}} in 2010. \citet{Pulla_DQSurvey_2016} published a revised version of \cite{Pushkarev_DQSurvey_2010} in 2016, with an extended list of 10 open-source DQ tools. All three surveys differ from our work since they do not map low-level functionalities of the tools to DQ dimensions. Also, the surveys are several years old and to not reflect recent developments in DQ tools. 

The closest work to ours is a large-scaled, systematic survey, conducted by Ehrlinger \& W\"{o}{\ss}~\cite{survey_ehringer}, who investigated 8 commercial and 5 open-source ones. The authors requested fully functional trial licenses via customer support for each commercial tools and also evaluated the functionality of the open-source tools.
Among other contributions, they provide insights regarding the coverage of DQ metrics and dimensions supported by the examined tools. In contrast to \cite{survey_ehringer}, we focus on the investigation of \emph{open-source} tools only. This choice enables us to investigate the source code of the examined tools directly, uncovering the actual functionalities. Hence, in this survey we are able to investigate \emph{low-level} functionalities from a \emph{technical} perspective, while \citet{survey_ehringer} investigated the functionalities of DQ tools from a \emph{user} perspective.

Another difference is the absence of predefined functionality requirements in our survey. While \citet{survey_ehringer} propose a comprehensive catalog of \textit{expected} functionalities, focusing on data profiling and automated DQ monitoring, we deliberately start with no predefined requirements, which enables us to investigate the functionalities that are actually used in practice from an \emph{unbiased} perspective.  
Due to the time that has passed, we were also able to investigate newer tools,  
only two of them (MobyDQ and Apache Griffin) overlapping with~\cite{survey_ehringer}.
Note that also for these two tools, we investigated a newer version.

\citet{survey_ehringer} conclude that (especially commercial) DQ tools often do not use the concept of DQ dimensions and metrics as suggested by research, and when they do, it is mainly for grouping different data validation checks.
We build upon this observation by investigating this connection between concrete data validation checks and high-level DQ dimensions more closely, enabling further research in how DQ dimensions are implemented in DQ tools.

\subsection{Related research on DQ dimensions}
\label{section:related_work_DQ_dimensions}
\citet{WangStrong1996} were one of the first to argue that DQ is a multifaceted concept and presented the first classification of DQ dimensions and their hierarchical structure.
In the meantime, a lot of research on DQ dimensions, different definitions, and possible classifications have been proposed~\cite{classifying_poor_data, Scannapieco_2002,Cichy_2019,mohammed2024qualityAssessmentVision}. 
Despite this large body of knowledge and several standards that have been developed (e.g., ISO/IEC~25012~\cite{iso25012} or ISO/IEC~25024~\cite{ISO25024}), there is still no agreement on which DQ dimensions are essential to use for a specific DQ project~\cite{Coleman_2013}. \citet{mohammed_2024-DQ-challenges} even go so far as to claim that a new perspective on DQ assessment is needed, because of the fragmented landscape of DQ dimensions. 

There has also been attempts to provide a mapping between DQ dimensions and specific DQ problems. While \citet{Almutiry_2015} provide a domain-specific mapping for electronic health records, ~\citet{classifying_poor_data} map related research about DQ problems to DQ dimensions. 
In contrast, we start from widely used DQ tools and investigate data validation checks in form of low-level functionalities implemented. 

To the best of our knowledge, this is the first work that (i) compiles a uniform view on functionalities for data validation and (ii) maps them to DQ dimensions. 
We therefore believe that this research offers insights into the practical implementation of DQ dimensions with results that support a common vision on how to assess data quality. 

\section{Conclusion and future work}
\label{section:conclusion_future_work}
In this survey, we have investigated the low-level functionalities of \numTools open-source DQ tools, highlighted their implementation variants, and mapped them to the most closely related DQ dimensions.
Thus, in contrast to existing surveys that solely investigate the functionality of DQ tools, we go one step further and analyze the connection of these low-level functionalities with high-level DQ dimensions.
To this end, we highlight the multiple perspectives for viewing single functionalities as well as the different functionalities that contribute to each DQ dimension.
We believe that the presented results are of interest to  both practitioners and researchers, since they offer a unified view of the fragmented landscape of DQ checks, while providing insights into what aspects are involved in the assessment of DQ dimensions. 

Regarding future work, more research should be devoted to resolve the current inconsistent terminology of DQ dimensions (as also highlighted in~\cite{mohammed_2024-DQ-challenges}). A unified view on DQ assessment would enable the development of standardized data verification checks for DQ tools.
In addition, we observed that the vast majority of open-source tools perform DQ assessment on relational data and do not support other data formats such as data streams or graphs. Hence, we would like to investigate the suitability of existing data verification checks to data streams. Considering additional challenges such as time-related dependencies and distribution in data streams, we believe that the development of new methods will be required.

\bibliographystyle{ACM-Reference-Format}
\bibliography{sample}

\appendix
\section{Exact names of low-level functionalities}
\label{appendix:function_names}

{\ok In this appendix, we present the exact names under which the identified low-level functionalities are found in the examined tools. Note that these names correspond to the source code elements (e.g., functions or classes) one can use in a software development environment to achieve the desired functionality. \autoref{table_functionalities_names_a} \& \autoref{table_functionalities_names_b} provide a mapping from the low-level functionalities to the exact tool implementations.\footnote{ Some names are hyphenated to fit in the line width. In these cases, the actual name does not contain a hyphen.}}
\resetFunctionalityId
\begin{table}[tb!]
    \caption{The exact names of the source code elements (e.g., functions or classes) provided by the examined tools for each identified low-level functionality (part 1 of 2).}
    \label{table_functionalities_names_a}
    \scriptsize
    \def\arraystretch{1.2}
    \begin{tabular}{p{0.22\linewidth} p{0.72\linewidth}}
        \toprule
        \multicolumn{1}{c}{\textbf{Low-level functionality}}      & \multicolumn{1}{c}{\textbf{Relevant function names in tools}}                                                                                                                                                                                                                                                                                                                                                                                                                                                                                                                                                                                                                                                                                                                                                                                                                                                                                                                                                                  \\
        \midrule
        \nextFunctionalityId~ values fall within range or set     & dbt Core (\texttt{accepted\_values}, \texttt{not\_accepted\_values}, \texttt{accepted\_range}, \texttt{relationships}, \texttt{relationships\_where}), Deequ (\texttt{isContainedIn}), Evidently (\texttt{TestValueRange}, \texttt{TestShareOfOutRangeValues}, \texttt{TestNumberOfOutRangeValues}, \texttt{TestCategoryShare}, \texttt{TestCategoryCount}, \texttt{TestValueList}, \texttt{TestNumberOfOutListValues}, \texttt{TestShareOfOutListValues}), GX (\texttt{expect\_column\_distinct\_values\_to\_be\_in\_set}, \texttt{expect\_column\_most\_common\_value\_to\_be\_in\_set}, \texttt{expect\_column\_pair\_values\_to\_be\_in\_set}, \texttt{expect\_column\_values\_to\_be\_in\_set}, \texttt{expect\_column\_values\_to\_not\_be\_in\_set}), Soda Core (\texttt{reference})                                                                                                                                                                                                                                    \\
        \nextFunctionalityId~ string values are of right length   & GX (\texttt{expect\_column\_value\_lengths\_to\_be\_between}, \texttt{expect\_column\_value\_lengths\_to\_equal}), Soda Core (\texttt{avg\_length}, \texttt{min\_length}, \texttt{max\_length})                                                                                                                                                                                                                                                                                                                                                                                                                                                                                                                                                                                                                                                                                                                                                                                                                                \\
        \nextFunctionalityId~ values comply with regex            & Deequ (\texttt{hasPattern}), Evidently (\texttt{TestColumnRegExp}), GX (\texttt{expect\_column\_values\_to\_match\_like\_pattern}, \texttt{expect\_column\_values\_to\_match\_like\_pattern\_list}, \texttt{expect\_column\_values\_to\_match\_regex}, \texttt{expect\_column\_values\_to\_match\_regex\_list}, \texttt{expect\_column\_values\_to\_not\_match\_like\_pattern}, \texttt{expect\_column\_values\_to\_not\_match\_like\_pattern\_list}, \texttt{expect\_column\_values\_to\_not\_match\_regex}, \texttt{expect\_column\_values\_to\_not\_match\_regex\_list}), Soda Core(\texttt{invalid\_count}, \texttt{invalid\_percent})                                                                                                                                                                                                                                                                                                                                                                                     \\
        \nextFunctionalityId~ timestamp values are recent         & dbt Core (\texttt{recency}), MobyDQ (\texttt{freshness}, \texttt{latency}), Soda Core (\texttt{freshness})                                                                                                                                                                                                                                                                                                                                                                                                                                                                                                                                                                                                                                                                                                                                                                                                                                                                                                                     \\
        \midrule
        \nextFunctionalityId~ row values satisfy comparison       & GX (\texttt{expect\_column\_pair\_values\_a\_to\_be\_greater\_than\_b}, \texttt{expect\_column\_pair\_values\_to\_be\_equal})                                                                                                                                                                                                                                                                                                                                                                                                                                                                                                                                                                                                                                                                                                                                                                                                                                                                                                  \\
        \nextFunctionalityId~ row satisfies SQL expression        & dbt Core (\texttt{expression\_is\_true}), Deequ (\texttt{satisfies}), Griffin (\texttt{SparkSQL}), MobyDQ (\texttt{validity})                                                                                                                                                                                                                                                                                                                                                                                                                                                                                                                                                                                                                                                                                                                                                                                                                                                                                                  \\
        \midrule
        \nextFunctionalityId~ simple descriptive statistic checks & Deequ (\texttt{Maximum}, \texttt{Minimum}, \texttt{Mean}, \texttt{Sum}, \texttt{hasMin}, \texttt{hasMax}, \texttt{hasMean}, \texttt{hasSum}, \texttt{hasStandardDeviation}, \texttt{hasEntropy}), Evidently (\texttt{TestColumnValueMin}, \texttt{TestColumnValueMax}, \texttt{TestColumnValueMean}, \texttt{TestColumnValueMedian}, \texttt{TestColumnValueStd}), GX (\texttt{expect\_column\_max\_to\_be\_between}, \texttt{expect\_column\_mean\_to\_be\_between}, \texttt{expect\_column\_median\_to\_be\_between}, \texttt{expect\_column\_min\_to\_be\_between}, \texttt{expect\_column\_stdev\_to\_be\_between}, \texttt{expect\_column\_sum\_to\_be\_between}, \texttt{expect\_column\_value\_z\_scores\_to\_be\_less\_than}, \texttt{expect\_column\_values\_to\_be\_between}, \texttt{expect\_multicolumn\_sum\_to\_equal}), Soda Core (\texttt{avg}, \texttt{max}, \texttt{min}, \texttt{sum}, \texttt{stdev}, \texttt{stdev\_pop}, \texttt{stdev\_samp}, \texttt{variance}, \texttt{var\_pop}, \texttt{var\_samp}) \\
        \nextFunctionalityId~ values satisfy ordering             & dbt Core (\texttt{sequential\_values}), GX (\texttt{expect\_column\_values\_to\_be\_increasing}, \texttt{expect\_column\_values\_to\_be\_decreasing})                                                                                                                                                                                                                                                                                                                                                                                                                                                                                                                                                                                                                                                                                                                                                                                                                                                                          \\
        \midrule
        \nextFunctionalityId~ columns match schema                & GX (\texttt{expect\_column\_to\_exist}, \texttt{expect\_table\_columns\_to\_match\_ordered\_list}, \texttt{expect\_table\_columns\_to\_match\_set}), Soda Core (\texttt{schema})                                                                                                                                                                                                                                                                                                                                                                                                                                                                                                                                                                                                                                                                                                                                                                                                                                               \\
        \nextFunctionalityId~ data types match schema             & Deequ (\texttt{DataType}), Evidently (\texttt{TestColumnsType}), GX (\texttt{expect\_column\_values\_to\_be\_dateutil\_parseable}, \texttt{expect\_column\_values\_to\_be\_of\_type}, \texttt{expect\_column\_values\_to\_be\_in\_type\_list}, \texttt{expect\_column\_values\_to\_be\_json\_parseable}, \texttt{expect\_column\_values\_to\_match\_json\_schema}, \texttt{expect\_column\_values\_to\_match\_strftime\_format}), Griffin (\texttt{schema\_conformance}), Soda Core (\texttt{schema})                                                                                                                                                                                                                                                                                                                                                                                                                                                                                                                          \\
        \midrule
        \nextFunctionalityId~ matching between tables             & dbt Core (\texttt{equality}), Deequ (\texttt{doesDatasetMatch}), Griffin (\texttt{accuracy})                                                                                                                                                                                                                                                                                                                                                                                                                                                                                                                                                                                                                                                                                                                                                                                                                                                                                                                                   \\
        \nextFunctionalityId~ adjacent intervals do not overlap   & dbt Core (\texttt{mutually\_exclusive\_ranges})                                                                                                                                                                                                                                                                                                                                                                                                                                                                                                                                                                                                                                                                                                                                                                                                                                                                                                                                                                                \\
        \bottomrule
    \end{tabular}
\end{table}

\begin{table}[tb!]
    \caption{The exact names of the source code elements (e.g., functions or classes) provided by the examined tools for each identified low-level functionality (part 2 of 2).}
    \label{table_functionalities_names_b}
    \scriptsize
    \def\arraystretch{1.2}
    \begin{tabular}{p{0.22\linewidth} p{0.72\linewidth}}
        \toprule
        \multicolumn{1}{c}{\textbf{Low-level functionality}}       & \multicolumn{1}{c}{\textbf{Relevant function names in tools}}                                                                                                                                                                                                                                                                                                                                                                                                                                                                                                                                                                                                                                                                                                                                                                                                       \\
        \midrule
        \nextFunctionalityId~ histogram checks                     & Deequ (\texttt{Histogram})                                                                                                                                                                                                                                                                                                                                                                                                                                                                                                                                                                                                                                                                                                                                                                                                                                          \\
        \nextFunctionalityId~ quantiles checks                     & Deequ (\texttt{hasExactQuantile}, \texttt{hasApproxQuantile}, \texttt{ApproxQuantiles}, \texttt{kllSketchSatisfies}), Evidently (\texttt{TestColumnQuantile}), GX (\texttt{expect\_column\_quantile\_values\_to\_be\_between}), Soda Core (\texttt{percentile})                                                                                                                                                                                                                                                                                                                                                                                                                                                                                                                                                                                                     \\
        \midrule
        \nextFunctionalityId~ checks on the number of elements     & dbt Core (\texttt{at\_least\_one}, \texttt{equal\_rowcount}, \texttt{fewer\_rows\_than}), Deequ (\texttt{hasSize}), Evidently (\texttt{TestNumberOfRows}, \texttt{TestNumberOfColumns}), GX (\texttt{expect\_table\_column\_count\_to\_be\_between}, \texttt{expect\_table\_column\_count\_to\_equal}, \texttt{expect\_table\_row\_count\_to\_be\_between}, \texttt{expect\_table\_row\_count\_to\_equal}, \texttt{expect\_table\_row\_count\_to\_equal\_other\_table}), Soda Core (\texttt{row\_count}, \texttt{cross})                                                                                                                                                                                                                                                                                                                                            \\
        \nextFunctionalityId~ distinct elements checks             & dbt Core (\texttt{cardinality\_equality}), Deequ (\texttt{hasNumberOfDistinctValues}, \texttt{hasApproxCount Distinct}, \texttt{hasDistinctness}), GX (\texttt{expect\_column\_distinct\_values\_to\_contain\_set}, \texttt{expect\_column\_distinct\_values\_to\_equal\_set})                                                                                                                                                                                                                                                                                                                                                                                                                                                                                                                                                                                      \\
        \nextFunctionalityId~ unique elements checks               & dbt Core(\texttt{unique}, \texttt{unique\_combination\_of\_columns}), Deequ (\texttt{hasUniqueness}, \texttt{isUnique}, \texttt{isPrimaryKey}, \texttt{hasUniqueValueRatio}), Evidently (\texttt{TestColumnAllUniqueValues}, \texttt{TestNumberOfUniqueValues}, \texttt{TestUniqueValuesShare}, \texttt{TestNumberOfDuplicatedRows}, \texttt{TestNumberOfDuplicatedColumns}), GX (\texttt{expect\_column\_proportion\_of\_unique\_values\_to\_be\_between}, \texttt{expect\_column\_unique\_value\_count\_to\_be\_between}, \texttt{expect\_column\_values\_to\_be\_unique}, \texttt{expect\_compound\_columns\_to\_be\_unique}, \texttt{expect\_multicolumn\_values\_to\_be\_unique}, \texttt{expect\_select\_column\_values\_to\_be\_unique\_within\_record}), Griffin (\texttt{duplication}), Soda Core (\texttt{duplicate\_count}, \texttt{duplicate\_percent}) \\
        \nextFunctionalityId~ most common elements checks          & dbt Core (\texttt{not\_constant}),  Evidently (\texttt{TestColumnAllConstantValues}, \texttt{TestNumberOfConstantColumns}, \texttt{TestMostCommonValueShare})                                                                                                                                                                                                                                                                                                                                                                                                                                                                                                                                                                                                                                                                                                       \\
        \nextFunctionalityId~ missing elements checks              & dbt Core (\texttt{not\_null}, \texttt{not\_empty\_string}, \texttt{not\_null\_proportion}), Deequ (\texttt{hasCompleteness, isComplete, areComplete, areAnyComplete, haveAnyCompleteness}), Evidently (\texttt{TestNumberOfMissingValues}, \texttt{TestShareOfMissingValues}, \texttt{TestNumberOfColumnsWithMissingValues}, \texttt{TestShareOfColumnsWithMissingValues}, \texttt{TestNumberOfRowsWithMissingValues}, \texttt{TestShareOfRowsWithMissingValues}, \texttt{TestColumnNumberOfMissingValues}, \texttt{TestColumnShareOfMissingValues}, \texttt{TestNumberOfEmptyRows}, \texttt{TestNumberOfEmptyColumns}), GX (\texttt{expect\_column\_values\_to\_be\_null}, \texttt{expect\_column\_values\_to\_not\_be\_null}), Griffin (\texttt{completeness}), Soda Core (\texttt{missing\_count}, \texttt{missing\_percent})                                    \\
        \nextFunctionalityId~ checks on represent. for missing     & Evidently (\texttt{TestNumberOfDifferentMissingValues}, \texttt{TestColumnNumberOfDifferentMissingValues})                                                                                                                                                                                                                                                                                                                                                                                                                                                                                                                                                                                                                                                                                                                                                          \\
        \midrule
        \nextFunctionalityId~ correlation checks                   & Deequ (\texttt{hasCorrelation}), Evidently (\texttt{TestTargetPredictionCorrelation}, \texttt{TestHighlyCorrelatedColumns}, \texttt{TestTargetFeaturesCorrelations}, \texttt{TestPredictionFeaturesCorrelations}, \texttt{TestCorrelationChanges})                                                                                                                                                                                                                                                                                                                                                                                                                                                                                                                                                                                                                  \\
        \nextFunctionalityId~ mutual information checks            & Deequ (\texttt{hasMutualInformation})                                                                                                                                                                                                                                                                                                                                                                                                                                                                                                                                                                                                                                                                                                                                                                                                                               \\
        \midrule
        \nextFunctionalityId~ anomaly detection                    & Deequ (\texttt{isNewestPointNonAnomalous})                                                                                                                                                                                                                                                                                                                                                                                                                                                                                                                                                                                                                                                                                                                                                                                                                          \\
        \nextFunctionalityId~ target labels/values do not conflict & Evidently (\texttt{TestConflictTarget}, \texttt{TestConflictPrediction})                                                                                                                                                                                                                                                                                                                                                                                                                                                                                                                                                                                                                                                                                                                                                                                            \\
        \nextFunctionalityId~ data drift detection                 & Evidently (\texttt{TestMeanInNSigmas}, \texttt{TestNumberOfDriftedColumns}, \texttt{TestShareOfDriftedColumns}, \texttt{TestColumnDrift}, \texttt{TestEmbeddingsDrift}), GX (\texttt{expect\_column\_kl\_divergence\_to\_be\_less\_than}), Soda Core (\texttt{distribution})                                                                                                                                                                                                                                                                                                                                                                                                                                                                                                                                                                                        \\
        \bottomrule
    \end{tabular}
\end{table}

\section{ISO/IEC 25012 Dimensions}
\label{appendix:dimensions}

We list here more details about the conceptual DQ model defined by the ISO/IEC 25012 Standard~\cite{iso25012}. The Standard defines fifteen data quality characteristics (i.e., dimensions). The textual descriptions of all characteristics are presented in \autoref{table_data_quality_dimensions}. The table comprises three parts, indicated by horizontal rules, namely \emph{Inherent} characteristics (top part), \emph{System-Dependent} characteristics (bottom part) and characteristics that belong to both categories (middle part). \emph{Inherent} data quality refers to the degree to which quality characteristics of data have the intrinsic potential to satisfy stated and implied needs when data is used under specified conditions. \emph{System-Dependent} data quality refers to the degree to which data quality is reached and preserved within a computer system when data is used under specified conditions. We refer the interested reader to~\cite{iso25012} for a more thorough analysis of the described DQ model. Lastly, we note that, apart from this standard, the literature is rich in theoretical DQ models~\cite[p. 3-5]{classifying_poor_data}, \cite{mohammed_glossary_2024}.

\begin{table*}[tb!]
    \caption{The DQ dimensions as defined by ISO/IEC 25012~\cite{iso25012}.}
    \label{table_data_quality_dimensions}
    \footnotesize
    \functionalitiesTablesStretch
    \begin{tabular}{p{0.13\linewidth} p{0.82\linewidth}} 
        \toprule
        \multicolumn{1}{l}{\textbf{Dimension}} & \multicolumn{1}{l}{\textbf{Definition}}                                                                                                                                                                                                                                                             \\
        \midrule
        Accuracy                               & The degree to which data has attributes that correctly represent the true value of the intended attribute of a concept or event in a specific context of use. It has two main aspects, namely \emph{Syntactic} and \emph{Semantic} accuracy.                                                        \\
        Completeness                           & The degree to which subject data associated with an entity has values for all expected attributes and related entity instances in a specific context of use.                                                                                                                                        \\
        Consistency                            & The degree to which data has attributes that are free from contradiction and are coherent with other data in a specific context of use. It can be either or both among data regarding one entity and across similar data for comparable entities.                                                   \\
        Credibility                            & The degree to which data has attributes that are regarded as true and believable by users in a specific context of use. Credibility includes the concept of authenticity (the truthfulness of origins, attributions, commitments).                                                                  \\
        Currentness                            & The degree to which data has attributes that are of the right age in a specific context of use.                                                                                                                                                                                                     \\
        \midrule
        Accessibility                          & The degree to which data can be accessed in a specific context of use, particularly by people who need supporting technology or special configuration because of some disability.                                                                                                                   \\
        Compliance                             & The degree to which data has attributes that adhere to standards, conventions or regulations in force and similar rules relating to data quality in a specific context of use.                                                                                                                      \\
        Confidentiality                        & The degree to which data has attributes that ensure that it is only accessible and interpretable by authorized users in a specific context of use. Confidentiality is an aspect of information security (together with availability, integrity) as defined in ISO/IEC 13335-1:2004~\cite{iso13335}. \\
        Efficiency                             & The degree to which data has attributes that can be processed and provide the expected levels of performance by using the appropriate amounts and types of resources in a specific context of use.                                                                                                  \\
        Precision                              & The degree to which data has attributes that are exact or that provide discrimination in a specific context of use.                                                                                                                                                                                 \\
        Traceability                           & The degree to which data has attributes that provide an audit trail of access to the data and of any changes made to the data in a specific context of use.                                                                                                                                         \\
        Understandability                      & The degree to which data has attributes that enable it to be read and interpreted by users, and are expressed in appropriate languages, symbols and units in a specific context of use. Some information about data understandability are provided by metadata.                                     \\
        \midrule
        Availability                           & The degree to which data has attributes that enable it to be retrieved by authorized users and/or applications in a specific context of use.                                                                                                                                                        \\
        Portability                            & The degree to which data has attributes that enable it to be installed, replaced or moved from one system to another preserving the existing quality in a specific context of use.                                                                                                                  \\
        Recoverability                         & The degree to which data has attributes that enable it to maintain and preserve a specified level of operations and quality, even in the event of failure, in a specific context of use.                                                                                                            \\

        \bottomrule
    \end{tabular}
\end{table*}

\end{document}